\documentclass[a4paper,11pt]{article}
\usepackage{jinstpub}

\usepackage{upgreek}
\usepackage{subfig}
\usepackage{xspace}
\usepackage[thinc]{esdiff}

\newcommand{\VGS}{\ensuremath{\textrm{V}_\textrm{gs}}\xspace}
\newcommand{\VDS}{\ensuremath{\textrm{V}_\textrm{ds}}\xspace}
\newcommand{\Vback}{\ensuremath{\textrm{V}_\textrm{back}}\xspace}
\newcommand{\Vmback}{\ensuremath{|{\textrm{V}_\textrm{back}}|}\xspace}
\newcommand{\ISD}{\ensuremath{\textrm{I}_\textrm{sd}}\xspace}
\newcommand{\Vth}{\ensuremath{\textrm{V}_\textrm{th}}\xspace}
\newcommand{\Vh}{\ensuremath{\textrm{V}_\textrm{1/2}}\xspace}
\newcommand{\Cox}{\ensuremath{\textrm{C}_\textrm{ox}}\xspace}
\newcommand{\epSi}{\ensuremath{{\epsilon}_\textrm{Si}}\xspace}
\newcommand{\epSiO}{\ensuremath{{\epsilon}_{\textrm{Si}\textrm{O}_2}}\xspace}
\newcommand{\qzero}{\ensuremath{\text{q}_\text{0}}\xspace}
\newcommand{\WbyL}{\ensuremath{\frac{\text{W}}{\text{L}}}\xspace}
\newcommand{\SiO}{\ensuremath{\textrm{Si}\textrm{O}_\textrm{2}}\xspace}
\newcommand{\mum}{\ensuremath{\upmu\textrm{m}}\xspace}
\newcommand{\mue}{\ensuremath{{\mu}_\textrm{e}}\xspace}

\title{\boldmath Characterization of p-stop isolation implants in silicon sensors using MOSFET structures}

\author[a]{T.~Bergauer,}
\author[b,1]{S.~Chatterjee,\note{Corresponding author.}}
\author[a]{M.~Dragicevic,}
\author[c]{I.~Kopsalis,}
\author[a,d]{and V.~Kraus}

\affiliation[a]{Marietta Blau Institute for Particle Physics, Austrian Academy of Sciences, Dominikanerbastei 16, 1010 Vienna, Austria}
\affiliation[b]{Deutsches Elektronen-Synchrotron DESY, Notkestra$\beta$e 85, 22607 Hamburg, Germany}
\affiliation[c]{Department of Physics, National Technical University of Athens, 9 Iroon Polytechniou St., Zografou Campus, 15780, Athens, Greece}
\affiliation[d]{European Organization for Nuclear Research CERN, CH-1211 Geneva, Switzerland}

\emailAdd{suman.chatterjee@cern.ch}

\abstract{
Metal-oxide-semiconductor field-effect transistor (MOSFET) test structures are investigated to characterize p-stop isolation implants between n-type electrodes in p-type silicon sensors. The device transfer characteristics are measured as a function of the voltage applied to the backside to extract the threshold voltage, which quantifies inter-electrode isolation, and the field-dependent mobility parameters. We present a methodology to reconstruct depth-dependent doping profiles from the threshold voltage characteristics, accounting for the localized space-charge effects and electric-field screening induced by the p-stop implants. The study evaluates the sensitivity of this technique to various p-stop geometries and doping concentrations across different wafer types. 
The results demonstrate the potential of MOSFET-based structures as a non-destructive diagnostic for monitoring p-stop consistency and inter-electrode isolation properties in silicon detectors.
}


\begin{document}

\maketitle
\flushbottom

\section{Introduction}
\label{Intro}

Silicon (Si) sensors in various forms are widely used for tracking, timing, and calorimetry in detectors for particle physics experiments. Current large-scale projects, such as the upgrades of the ATLAS and CMS experiments for the High-Luminosity phase of the Large Hadron Collider (HL-LHC), incorporate hundreds of square meters of Si sensors and therefore require large-scale industrial production capabilities. To ensure quality and long-term stability of those production processes, rigorous quality control procedures based on dedicated test structures have been established, as described, for example, in Ref.~\cite{Hinger:2021jkr}.

A key feature of segmented Si sensors is the implementation of isolation implants, such as p-stop structures,  used in n-on-p (p-type) Si sensors.
These implants electrically insulate adjacent n-type charge collection electrodes and prevent them from being short-circuited by an accumulation layer of electrons that forms at the interface of Si and silicon dioxide (\SiO). 
While these implants are essential for device operation, they locally modify the electric field distribution and effective doping profile near the surface, thereby influencing device characteristics.
 
\begin{figure}[hbtp]
    \centering
    {\includegraphics[width=0.495\textwidth]{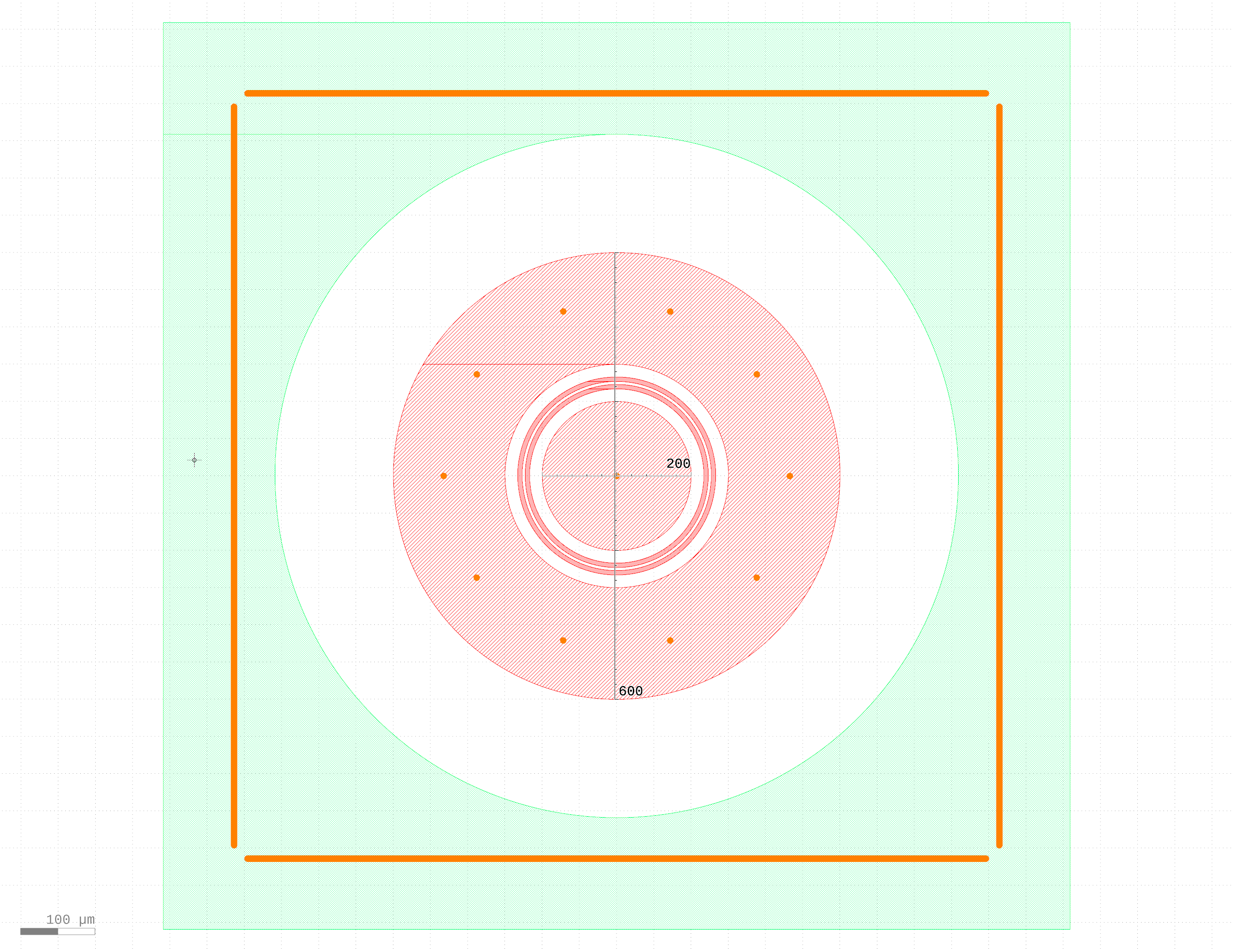}}
    {\includegraphics[width=0.495\textwidth]{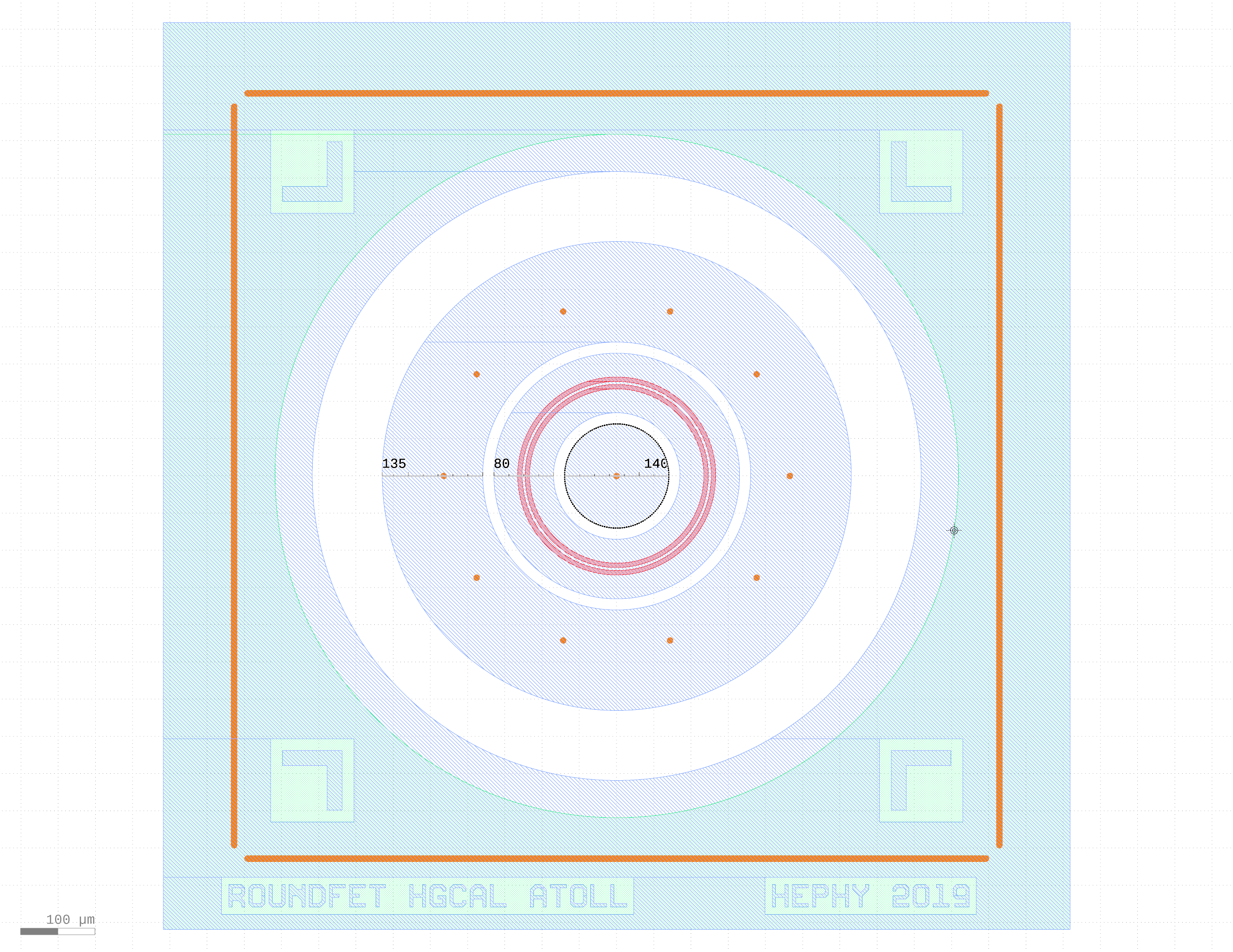}}
    {\includegraphics[width=0.45\textwidth]{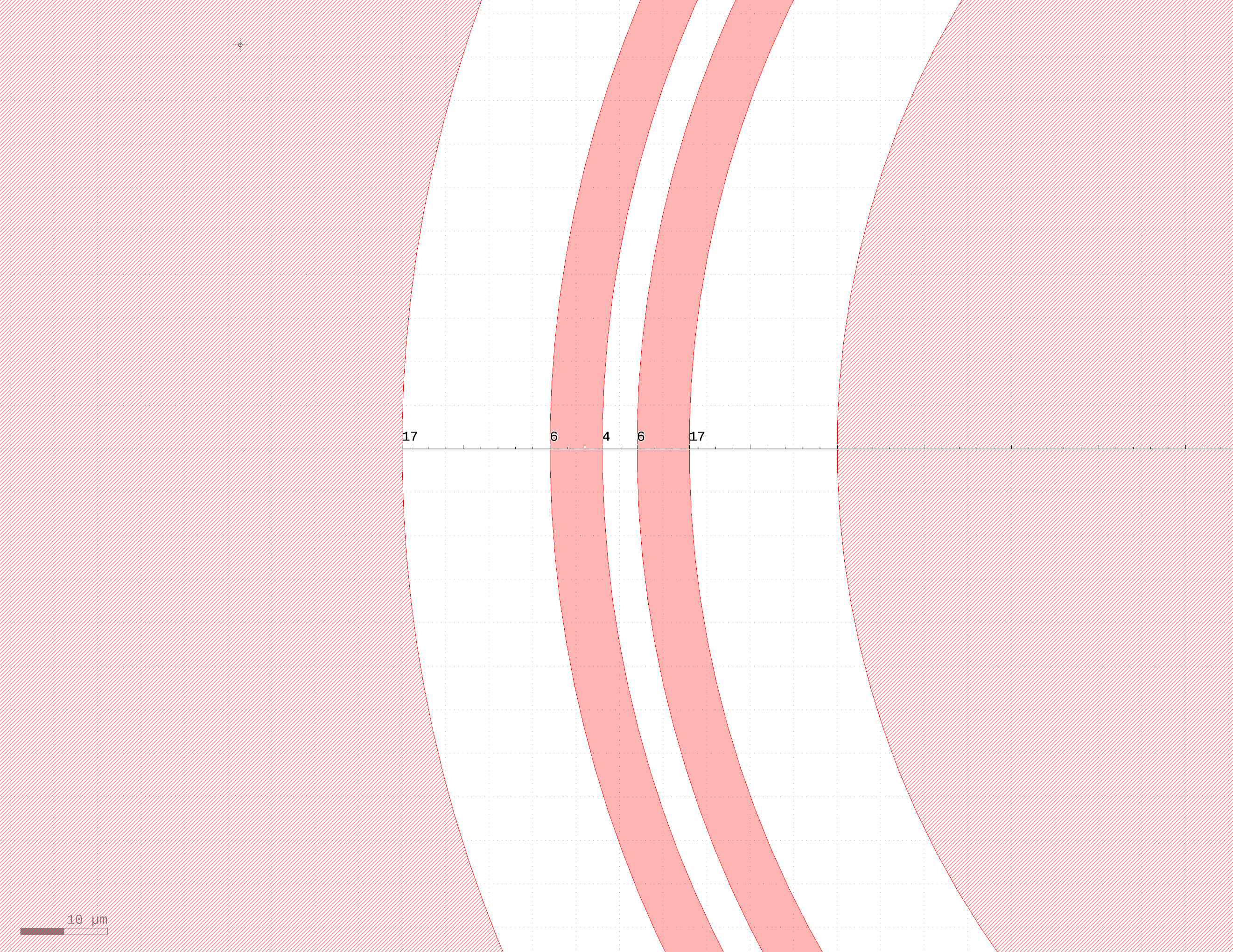}}
    {\includegraphics[width=0.45\textwidth]{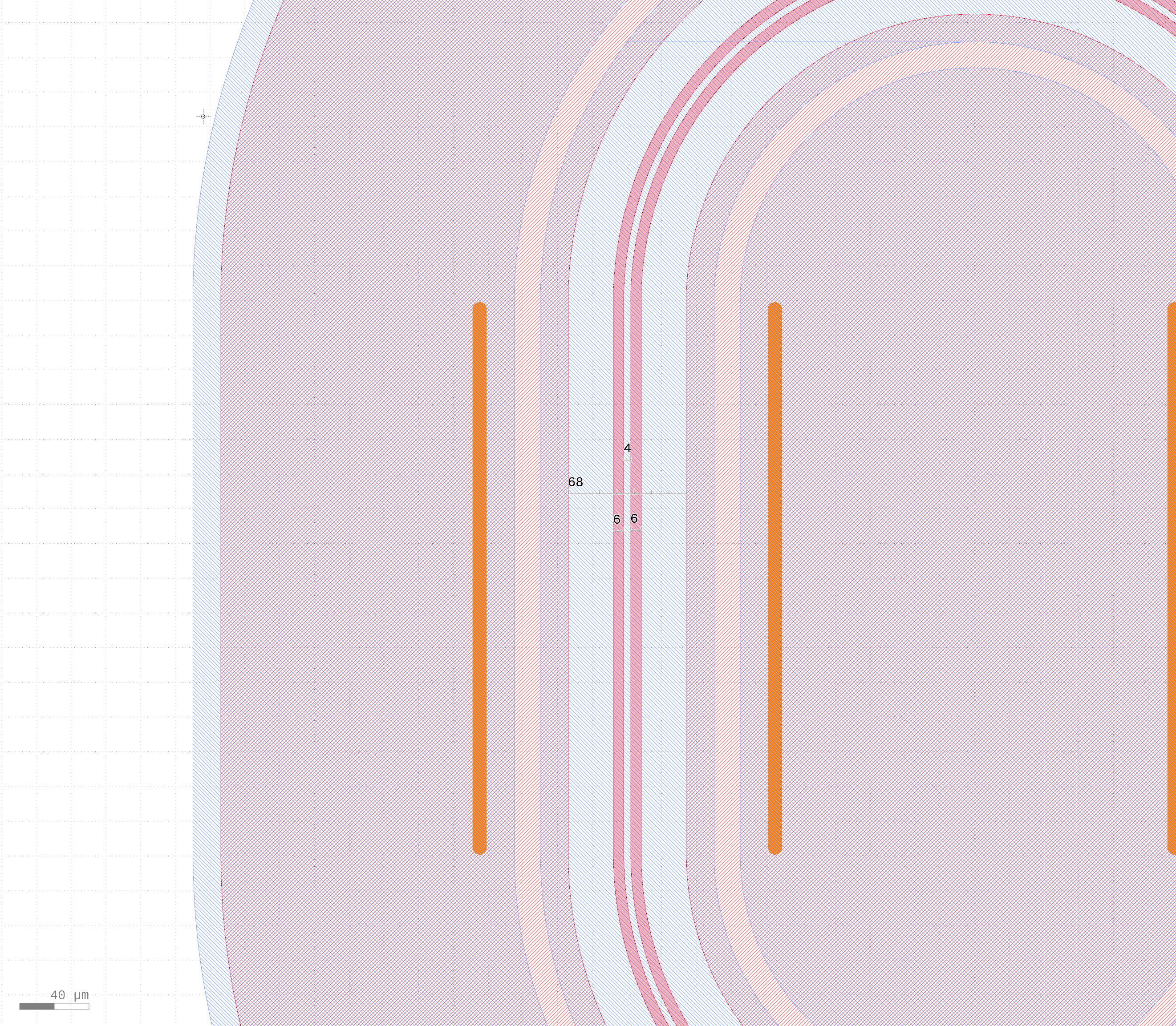}}
    \caption{
    GDS layout of an exemplary circular MOSFET test structure: (top left) highlighting the p-stop implants as narrow rings (red), (top right) showing the n$^{+}$ implants located in the white region and the aluminum metallization layer (blue). 
    The innermost circle defines the source electrode and the outer hatched ring with small orange contact pads represents the drain electrode of the MOSFET.
    Close-up views of the channel region, featuring two p-stop implants, are shown for the circular (bottom left) and elliptical (bottom right) geometries.
    }
    \label{fig:FET}
\end{figure}

To study the characteristics of p-stop implants in detail, we designed field-effect transistors (FETs) fabricated on wafers representative of silicon-sensor production. 
These devices are test structures designed to replicate the geometry of the inter-electrode region in segmented sensors, which is the area between neighboring strips, pixels, or pads in actual sensors. 
The present study serves as a proof of concept for using their electrical response as a non-destructive probe of p-stop implantation and inter-electrode isolation. 
In these structures, the adjacent collection electrodes function as the source and drain terminals, while the gate is placed atop the {\SiO} layer between them, as in a metal-oxide-semiconductor FET (MOSFET). 
To mimic various sensor designs, the channel between the electrodes can be configured without p-stops or with one or two p-stop implants, reflecting the isolation scheme used in actual sensors.
Two different geometries are employed, featuring concentric circular and elliptical arrangements of electrodes and p-stop implants. 
Examples of such test structures with two p-stop implants are illustrated in Fig.~\ref{fig:FET}. 

By grounding the source and applying an electric potential to the gate, the charge carrier density in the channel, and thus the current flow between source and drain, can be controlled. A fundamental parameter of MOSFET structures is the threshold voltage (\Vth), defined as the gate-to-source voltage (\VGS) at which the drain-to-source channel becomes conductive. 
The value of \Vth depends on several factors, primarily the bulk doping concentration and, whenever relevant, the doping profile of the p-stop implants~\cite{Konig_2017,HINGER2020162233}.

In this paper, we investigate the evolution of {\Vth} as a function of the backside bias voltage (\Vback).
The potential applied to the backside affects the depletion region formed beneath the channel at the Si-{\SiO} interface and thereby alters {\Vth}. 
We further leverage this variation and the associated change in depletion depth to extract information on the effective bulk doping and the p-stop implantation dose under appropriate assumptions. 
This approach provides a non-destructive method to characterize doping profiles, which is critical for simulating sensor characteristics and gauging their performance under diverse operating conditions.

In this work, we analyze circular and elliptical n-channel MOSFETs fabricated on wafers with different bulk doping concentrations and isolation implant configurations. Devices with and without p-stop implants are compared in order to disentangle bulk and surface-related effects. 
First, the electron mobility is extracted from transfer characteristics using a parameterized model for the mobility. 
Subsequently, the threshold voltage is determined as a function of backplane bias, i.e., the voltage applied to the MOSFET backside. 
From the back-bias dependence of the threshold voltage, the depletion depth is inferred and used to reconstruct the effective doping profile via an analytical relation.

The paper is organized as follows. 
Section~\ref{sec:setup} describes the measurement setup and specific MOSFET structures considered. 
The analysis of the measured data, including the methodology for extracting {\Vth} and determining doping parameters, is discussed in Sec.~\ref{sec:results}. 
Finally, we summarize the findings and conclude in Sec.~\ref{sec:conclusion}.

\section{Test structures and measurement setup}
\label{sec:setup}

Different MOSFET structures fabricated on p-type Si were used in the measurements.
%The devices replicate the inter-electrode region between two n$^{+}$ strips and 
The devices feature a gate electrode electrically isolated from the Si bulk and possible p-stop implants by a {\SiO} dielectric layer.

The MOSFETs are fabricated on two distinct wafer types, designated T and H, with diameters of 6 and 8 inches, respectively. 
Both wafer types have a thickness of approximately 300\;{\mum}.
These wafers have also been used to produce sensors for the Outer Tracker~\cite{ref:Tracker} and High-Granularity Calorimeter~\cite{ref:HGCAL} to be used in the CMS experiment at the HL-LHC.

Of the two MOSFET geometries studied, the circular MOSFETs are available on wafer T, whereas the elliptical MOSFET structures are implemented on both wafer types.
The two layouts are included to study whether the quantitative extraction of device parameters from the transfer characteristics remains applicable across different electrode geometries.
While the rotational symmetry of the circular geometry facilitates a simple geometric parameterization and minimizes edge effects, the elliptical geometry more closely replicates the elongated, straight inter-electrode regions in segmented detector layouts and is well suited to probe-based measurements.
The {\SiO} layer, located between the metal gate and the Si bulk, is 700--750\;nm thick on T and H wafers, as determined from oxide capacitance measurements of metal-oxide-semiconductor capacitors.
A summary of the MOSFET variants along with the naming convention adopted in this paper is provided in Table~\ref{tab:MOSFETs}. 

\begin{table}[hbtp]
\caption{
Naming convention used for the MOSFETs and a brief description of each structure.
}
\begin{center}
{\renewcommand{\arraystretch}{1.3}
\begin{tabular}{lcl|lcl}
Circular & Wafer & Description &  Elliptical & Wafer & Description \\
\hline
T-P1  & T & 1 p-stop   & H-PLa & H & 2 p-stops with low doping \\
T-P2  & T & 2 p-stops  & H-PLb & H & 2 p-stops with low doping \\
T-P0 						  & T & No p-stop  & H-PHa & H & 2 p-stops with high doping \\
  							  &  		 & 		      & H-PHb & H & 2 p-stops with high doping \\
  	 						  &  		 & 		      & T-PL  & T & 2 p-stops with low doping \\
  	 						  &  		 & 		      & H-P0  & H & No p-stop \\
%\hline
\end{tabular}}
\label{tab:MOSFETs}
\end{center}
\end{table}

The p-stop doping concentrations of T-P1 and T-P2 are comparable to those of H-PLa and H-PLb. In contrast, H-PHa and H-PHb have higher p-stop doping levels, as inferred from p-stop resistivity measurements of van der Pauw-type cross structures. 
Furthermore, the resistivity of H-PLb (H-PHb) is slightly lower compared to that of H-PLa (H-PHa).

\begin{figure}[hbtp]
    \centering
    {\includegraphics[width=0.95\textwidth]{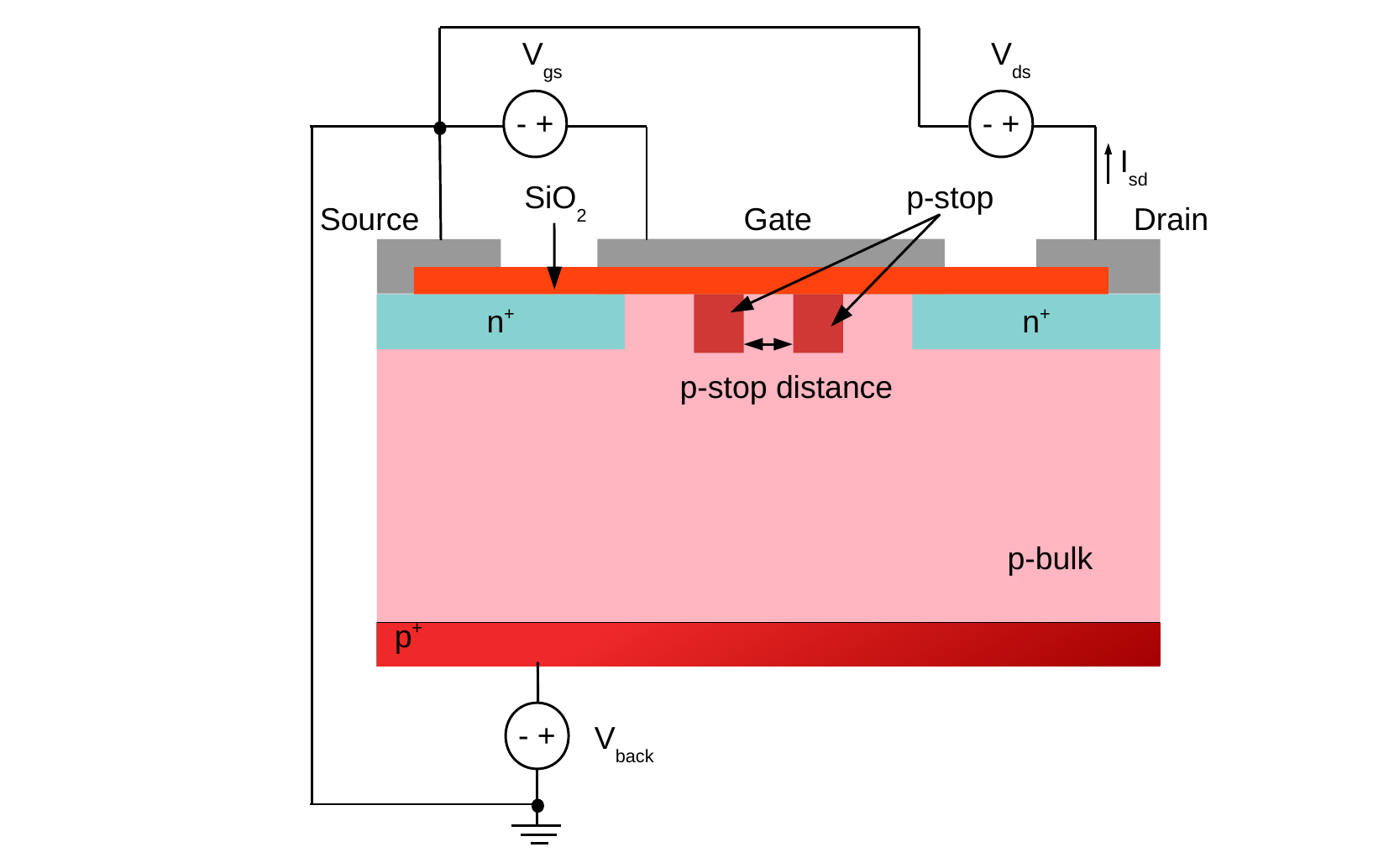}}
    \caption{
    Schematic cross section of a MOSFET test structure overlaid with the circuit diagram corresponding to the measurement configuration.
    }
    \label{fig:circuit}
\end{figure}

A schematic diagram of the measurement configuration is shown in Fig.~\ref{fig:circuit}. 
All measurements were conducted at room temperature under ambient humidity. 
The MOSFETs were mounted on a chuck inside a probe station, with the source electrode held at the ground potential.
The voltages {\Vback} and {\VGS}, as well as the drain-to-source voltage (\VDS), were supplied using three Keithley 2470 source measure units. 
The source-to-drain current (\ISD) was recorded with a Keithley 6517b electrometer. 

As p-n junctions are operated under reverse bias for particle detection in high-energy physics experiments, {\Vback} values, ranging from $0.5$\;V to $-300$\;V, were applied.
For each \Vback value, \VGS was varied from 0 to 35\;V in steps of $0.1$\;V and the corresponding {\ISD} values were recorded. 
The \VDS value was fixed at 50\;mV in each case.
For the two structures with high p-stop doping, H-PHa and H-PHb, the {\Vback} range was extended to $-600$\;V to ensure full depletion.
It was verified that the data obtained by ramping \VGS up and down were consistent.

\section{Data analysis and results}
\label{sec:results}

In the following sections, we describe the methodology used to extract device parameters from the analysis of experimental data. This study focuses on the determination of {\Vth} and the characterization of the p-stop implant doping profiles. 

\subsection{Estimation of threshold voltage and mobility parameters}
\label{sec:Vth}

\begin{figure}[hbtp]
    \centering
    {\includegraphics[width=0.99\textwidth]{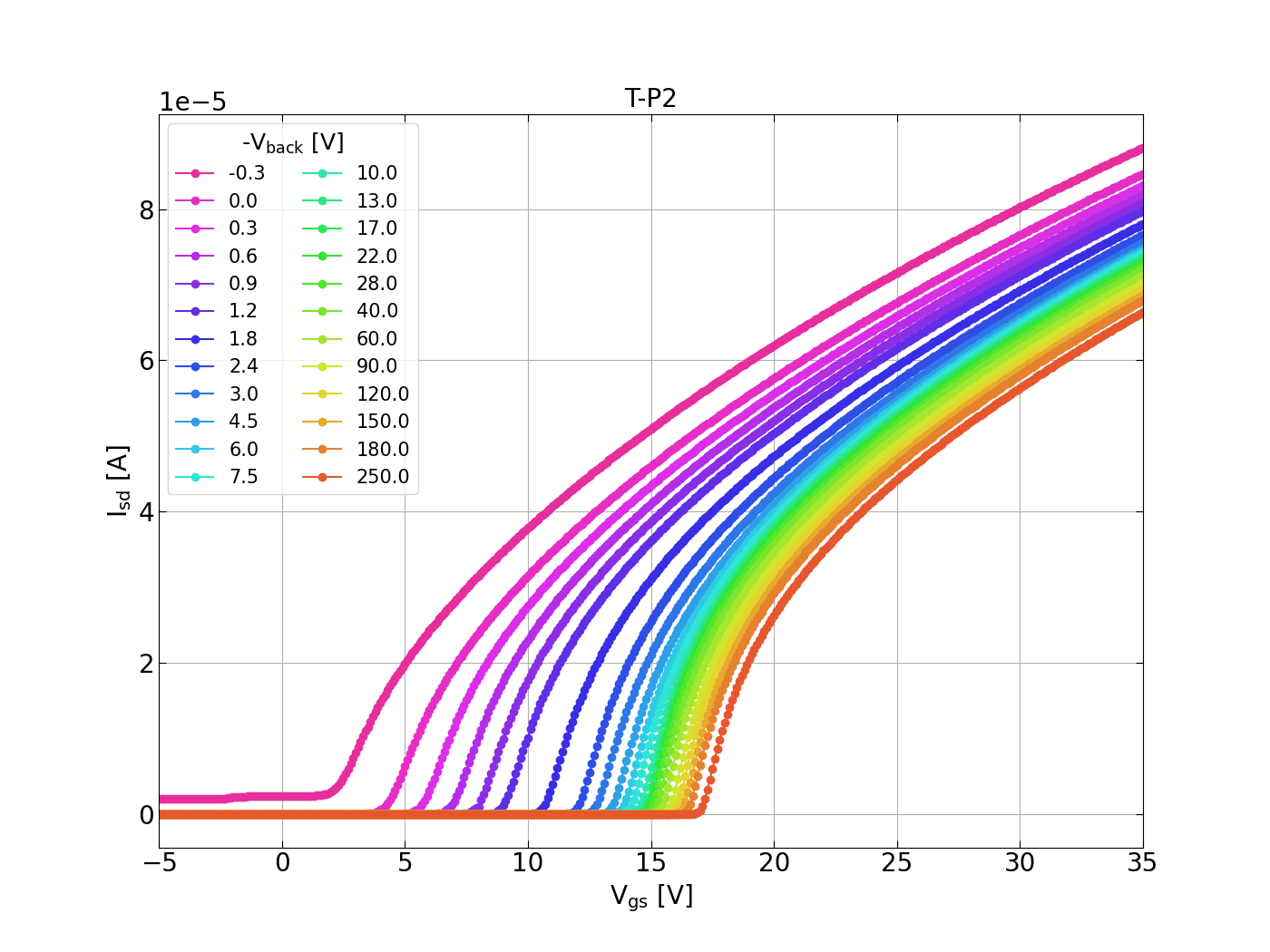}}
    {\includegraphics[width=0.99\textwidth]{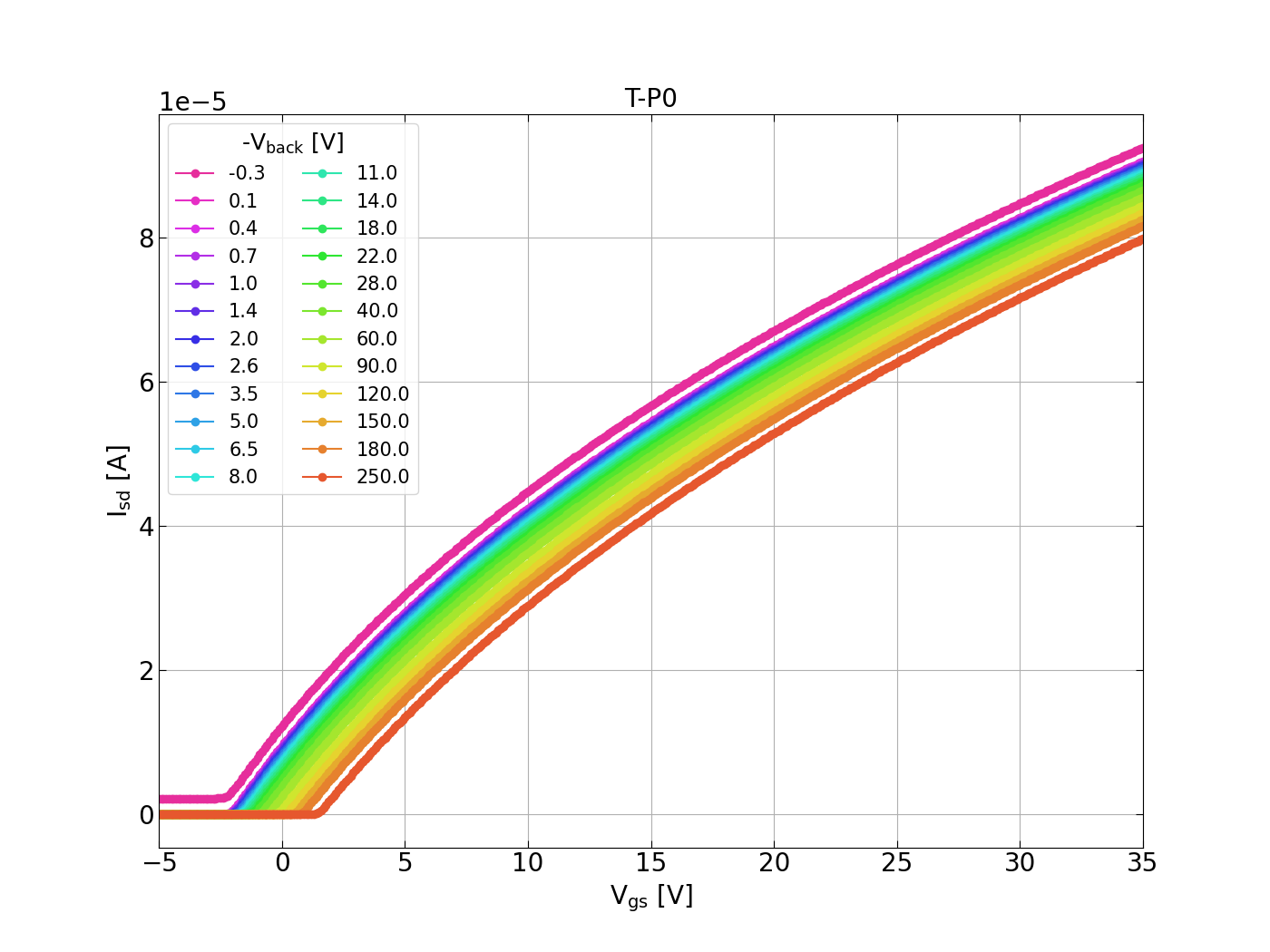}}
    \caption{
	Measured transfer characteristics, {\ISD}-vs-{\VGS}, for the T-P2 (top) and T-P0 (bottom) MOSFETs at selected {\Vback} values, indicated by different colors.
    }
    \label{fig:ISD_VGS}
\end{figure}

The variation of \ISD as a function of \VGS, representing the MOSFET transfer characteristics, is shown in Fig.~\ref{fig:ISD_VGS} for selected \Vback values for two structures on wafer T, T-P2 and T-P0, differing in the p-stop configuration as detailed in Table~\ref{tab:MOSFETs}. 

The general behavior, common to all \Vback values, is that \ISD remains close to zero until \VGS reaches {\Vth}, beyond which \ISD increases continuously.
The T-P2 structure exhibits a pronounced dependence of \Vth on {\Vback}, which is caused by the presence of p-stop implants.
In contrast, for the T-P0 structure, which does not have any p-stops, {\Vth} is negative unless {\Vmback} becomes sufficiently large, and the {\ISD}-vs-{\VGS} profiles exhibit a marginal shift for different {\Vback} values.
Apart from the near-zero bias regime ($\Vback \geq -0.3$\;V), the overall shape of the {\ISD}-vs-{\VGS} profiles remains similar across different \Vback values for both structures.
The same qualitative behavior is observed across all geometries and wafer types considered in the measurements.

\begin{figure}[hbtp]
    \centering
    {\includegraphics[width=0.495\textwidth]{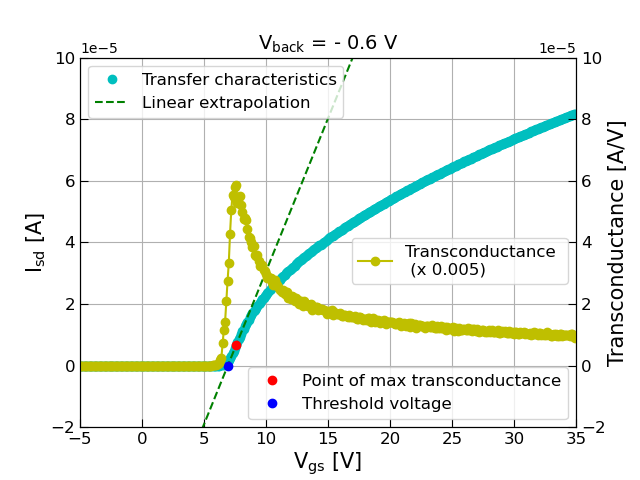}}
    {\includegraphics[width=0.495\textwidth]{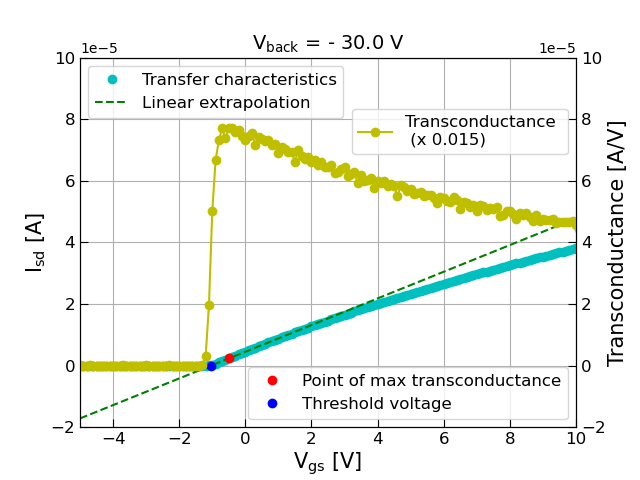}}
    \caption{
    Visual illustration of the ELR method for the T-P2 (left) and T-P0 (right) MOSFETs at two {\Vback} values. The x-axis range is zoomed in for T-P0 for better visibility. The transfer characteristics and transconductance are shown as cyan and yellow points, respectively, while the point of maximum transconductance is indicated by a red marker, and the threshold voltage is represented by a blue marker. Transconductance values are scaled to the range of the transfer characteristics.
    }
    \label{fig:transconductance}
\end{figure}

An initial estimate of \Vth is obtained using the extraction in the linear region (ELR) method. 
In this approach, the voltage corresponding to the maximum value of transconductance, defined as the first derivative of the transfer characteristics, is obtained. 
A linear extrapolation of the transfer characteristics  at this point is then performed, and the intercept of the extrapolated line on the \VGS axis is taken as {\Vth}. 
The procedure is illustrated in Fig.~\ref{fig:transconductance}.

In order to reduce the effects of measurement fluctuations on the estimate of {\Vth}, the {\ISD}-vs-{\VGS} data points are fitted with a mobility model~\cite{BREWS1978345} as expressed in Eq.~\eqref{Eq:mobility_model}.
\begin{equation}
{\ISD} = {\WbyL} {\mue}\xspace {\Cox} {\VDS} (\VGS - \Vth) \;
\label{Eq:mobility_model}
\end{equation}
For circular geometries, the effective width-to-length ratio, {\WbyL}, in Eq.~\eqref{Eq:mobility_model} is obtained using the relation ${\WbyL} = \frac{2\pi}{\ln\left(\frac{r_\text{D}}{r_\text{S}}\right)}$, where $r_\text{S}$ and $r_\text{D}$ are the outer radius of the source implant and the inner radius of the drain implant, respectively.
The oxide capacitance per unit area, $\Cox$, is determined from the oxide thickness ($\text{d}_\text{ox}$) via $\Cox = \frac{\epSiO}{\text{d}_\text{ox}}$, where {\epSiO} is the permittivity of {\SiO}. The parameter values are summarized in Table~\ref{tab:parameters}.

\begin{table}[hbtp]
\caption{
Geometric and dielectric parameters of the MOSFET structures on wafers T and H. 
}
\begin{center}
\begin{tabular}{c|c|c}
Parameter & \multicolumn{2}{c}{Value} \\
& T wafer & H wafer \\
\hline
$\text{d}_\text{ox}$ & 720 nm & 740 nm \\
\Cox &  4.8 nF/cm$^2$  & 4.7 nF/cm$^2$  \\
$r_\text{S}$ & 50\;{\mum} & 50\;{\mum}\\
$r_\text{D}$ & 75\;{\mum} & 75\;{\mum}\\
\end{tabular}
\label{tab:parameters}
\end{center}
\end{table}

In Eq.~\eqref{Eq:mobility_model}, {\mue} refers to the electron mobility, which is parameterized as a function of {\VGS} following Eq.~\eqref{Eq:mobility_mu}~\cite{schroder2015semiconductor}
\begin{equation}
{\mue} = {\mu}_0 \frac{1}{1+\frac{ \left(\VGS - \Vth \right)}{\Vh}} \, ,
\label{Eq:mobility_mu}
\end{equation}
where ${\mu}_0$ refers to the electron mobility at the Si-{\SiO} interface when \VGS just reaches {\Vth} and \Vh is the characteristic value of $\left(\VGS - \Vth\right)$ at which the electron mobility is reduced to half of its value at threshold due to surface scattering.

\begin{figure}[hbtp]
    \centering
    {\includegraphics[width=0.495\textwidth]{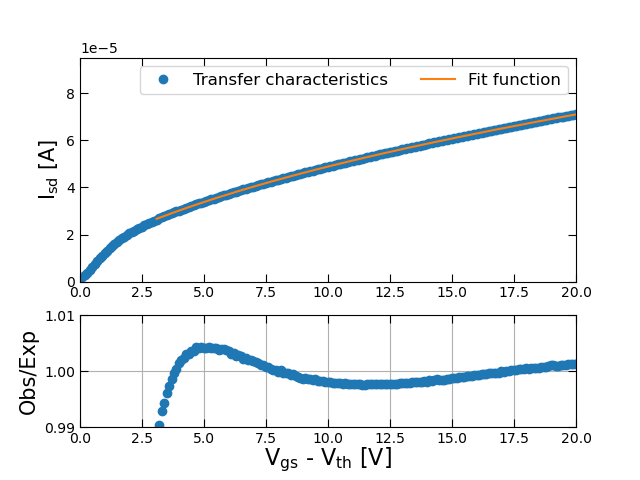}}
    {\includegraphics[width=0.495\textwidth]{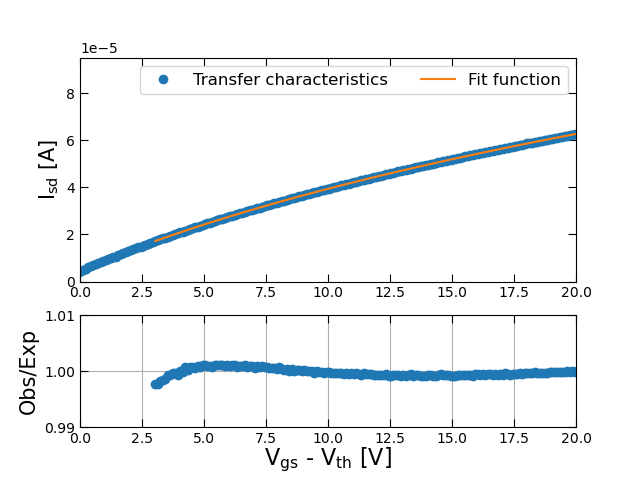}}
    \caption{
    Fits to the transfer characteristics (blue) using the function (orange) defined in Eq.~\eqref{Eq:mobility_model} for the T-P2 (left) and T-P0 (right) MOSFETs at $\Vback = 0$.
    }
    \label{fig:ISD_VGS_fit}
\end{figure}

Exemplary fits of the transfer characteristics with the mobility model are shown in Fig.~\ref{fig:ISD_VGS_fit} for the T-P2 and T-P0 structures, considering the case in which no backside voltage is applied, i.e., $\Vback = 0$. 
Only data points with $\left(\VGS - \Vth\right) \geq 3$\;V are included in the fit, as data below this range exhibit non-linear behavior not captured by the mobility model defined in Eq.~\eqref{Eq:mobility_model}.
As shown in the ratio panels of Fig.~\ref{fig:ISD_VGS_fit}, the model describes the data to within $0.5\%$ ($0.1\%$) for T-P2 (T-P0). 
The better agreement for T-P0 is attributed to the absence of p-stop implants, which, in T-P2, introduce additional non-linear effects in the near-threshold region.

\begin{figure}[!t]
    \centering
    {\includegraphics[width=0.495\textwidth]{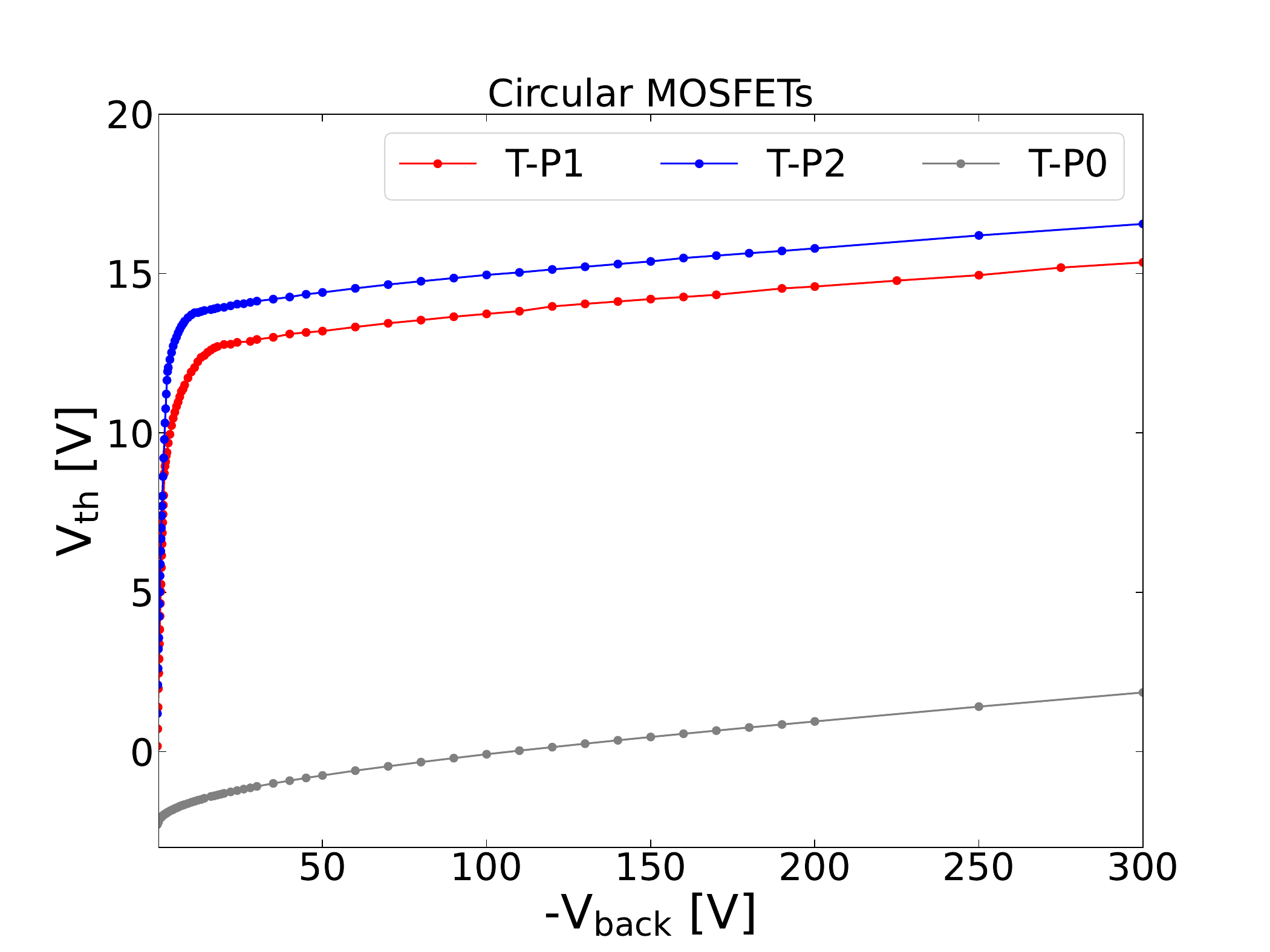}}
    {\includegraphics[width=0.495\textwidth]{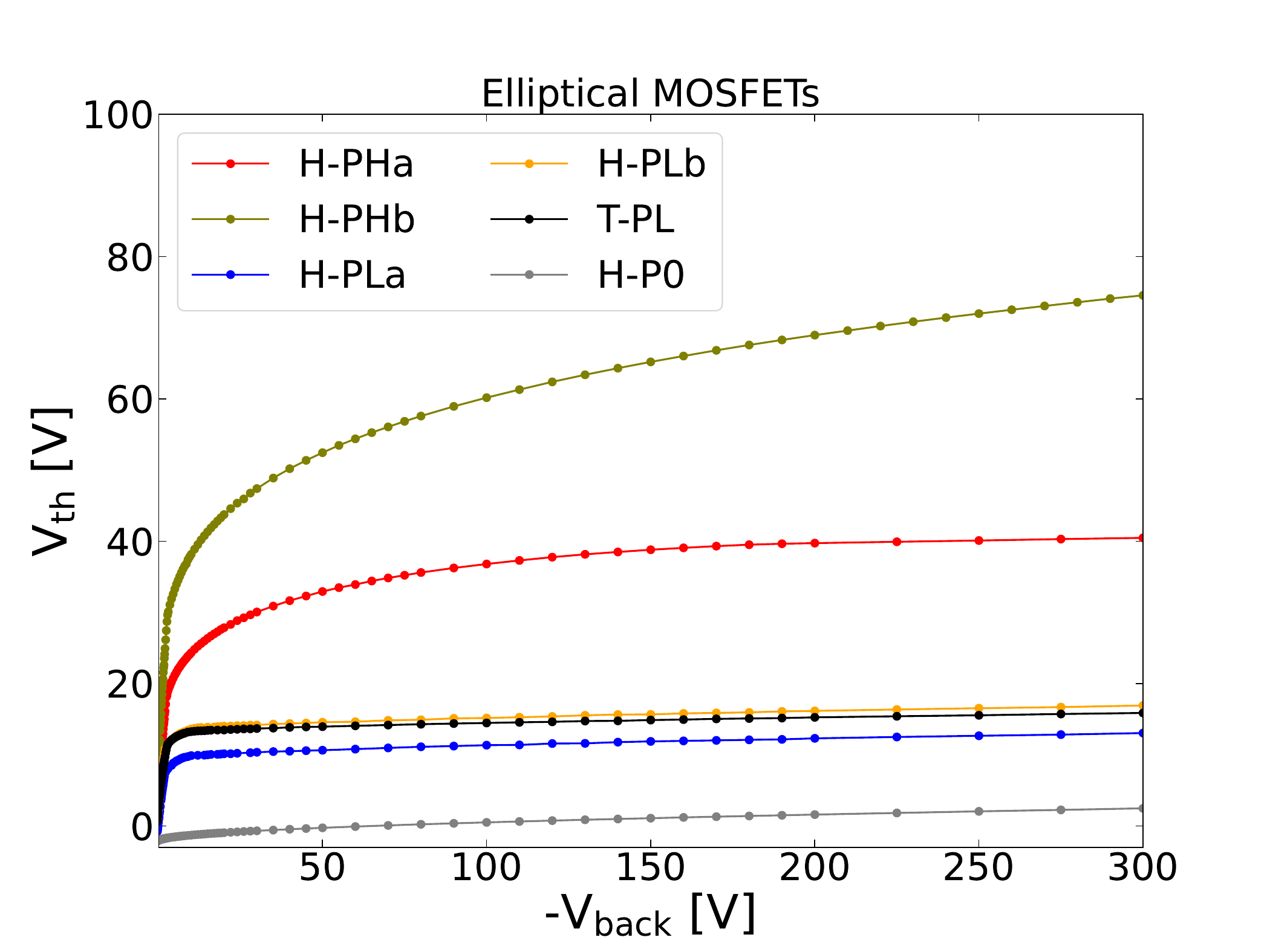}}
    \caption{
    Threshold voltage as a function of \Vback for circular (left) and elliptical (right) MOSFET structures. Each color corresponds to a different p-stop configuration for the respective MOSFET geometry.
    }
    \label{fig:VthvsVbias}
\end{figure}

Based on the combined model of Eqs.~\eqref{Eq:mobility_model} and~\eqref{Eq:mobility_mu}, the parameters  ${\mu}_0$, {\Vh}, and {\Vth} are extracted for each {\Vback} value.
The variation of the resulting \Vth with {\Vback} is shown in Fig.~\ref{fig:VthvsVbias} for circular and elliptical geometries.
The circular MOSFETs on wafer T with p-stop implants, T-P1 and T-P2, show very similar behavior, where \Vth initially increases steeply with {\Vmback}, followed by a knee region beyond which the rate of increase of \Vth becomes very small. 
A similar trend is observed for the elliptical structures with low p-stop doping (T-PL, H-PLa, H-PLb). In contrast, for the high-doping structures H-PHa and H-PHb, the region of sharp \Vth increase extends over a much larger {\Vmback} range, indicating a higher voltage requirement to reach a similar depletion state.
For the structures without p-stop implants, \Vth changes only slowly with \Vmback throughout the entire measurement range, and this behavior is independent of geometry. 
The uncertainty in {\Vth}, estimated using the covariance matrix of the fitted mobility-model parameters, is found to be in the range of 0.5--1\%. 
The contribution to the {\Vth} uncertainty arising from the resolution of the current and voltage measurements is at the sub-per-mille level and is therefore negligible in comparison.

\begin{figure}[!t]
    \centering
    {\includegraphics[width=0.495\textwidth]{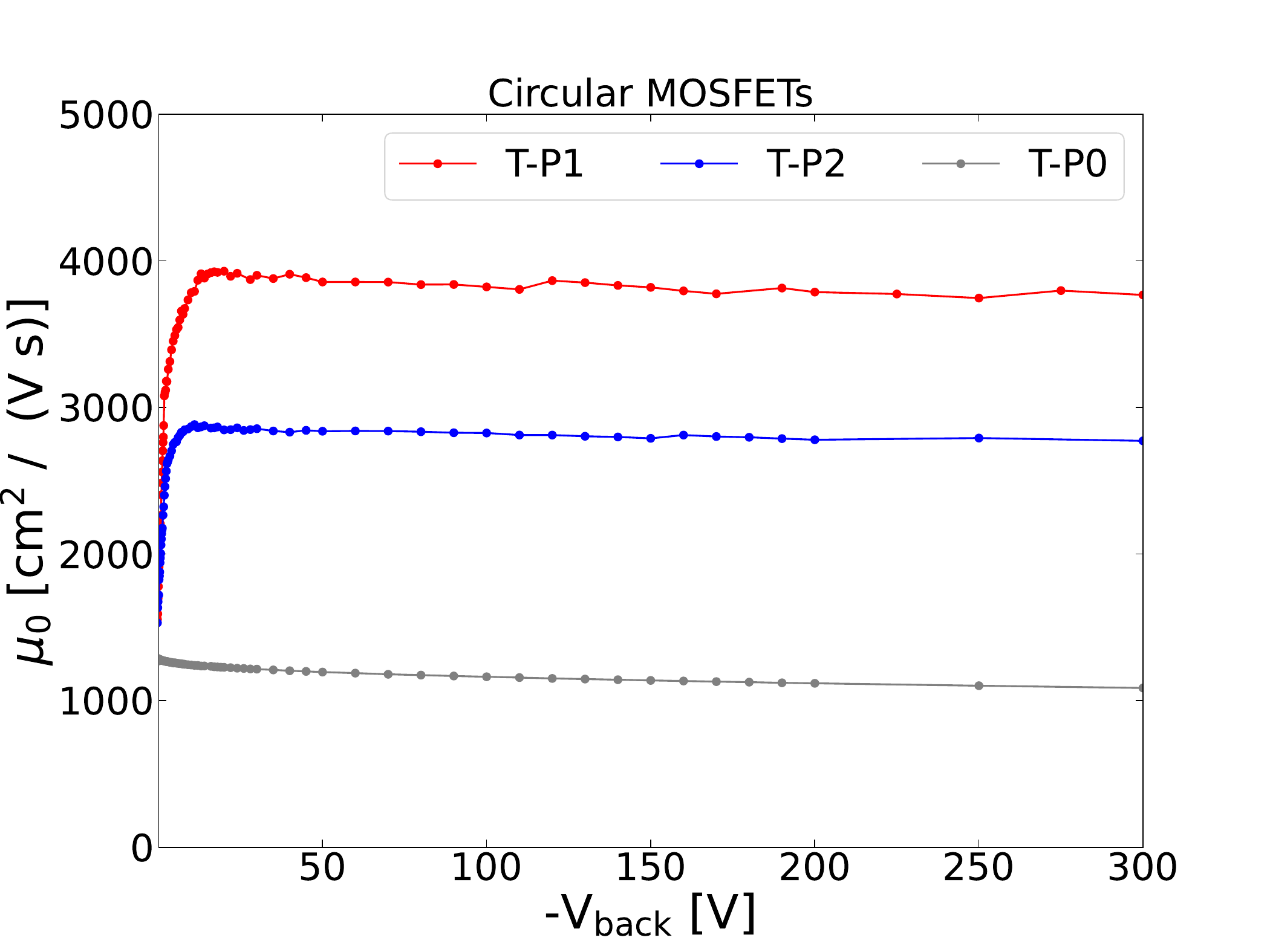}}
    {\includegraphics[width=0.495\textwidth]{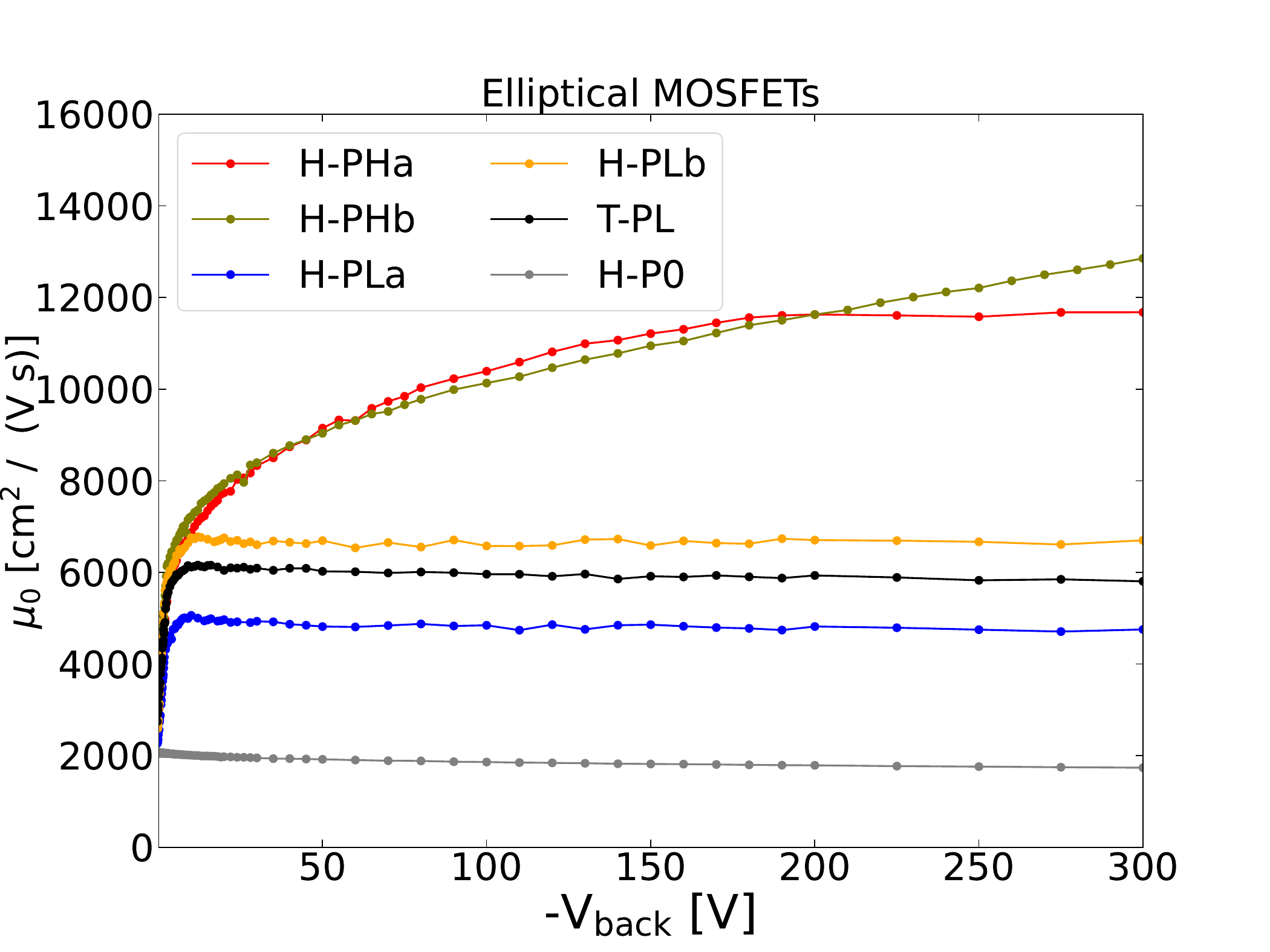}}
    \caption{
    Values of the ${\mu}_0$ parameter as a function of {\Vback} for circular (left) and elliptical (right) MOSFET structures. Each color corresponds to a different p-stop configuration for the respective MOSFET geometry.
    }
    \label{fig:mu0vsVbias}
\end{figure}

\begin{figure}[!t]
    \centering
    {\includegraphics[width=0.495\textwidth]{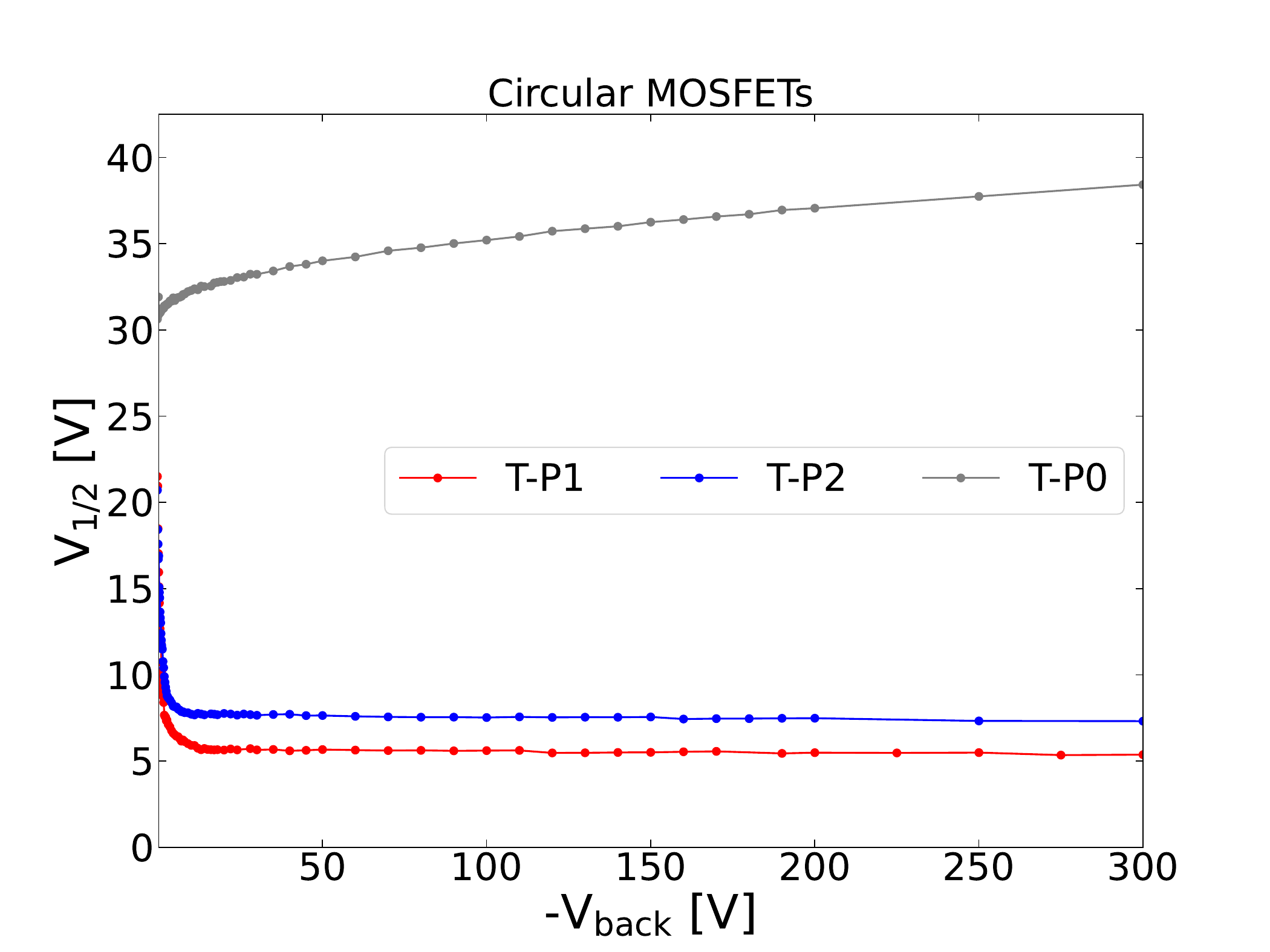}}
    {\includegraphics[width=0.495\textwidth]{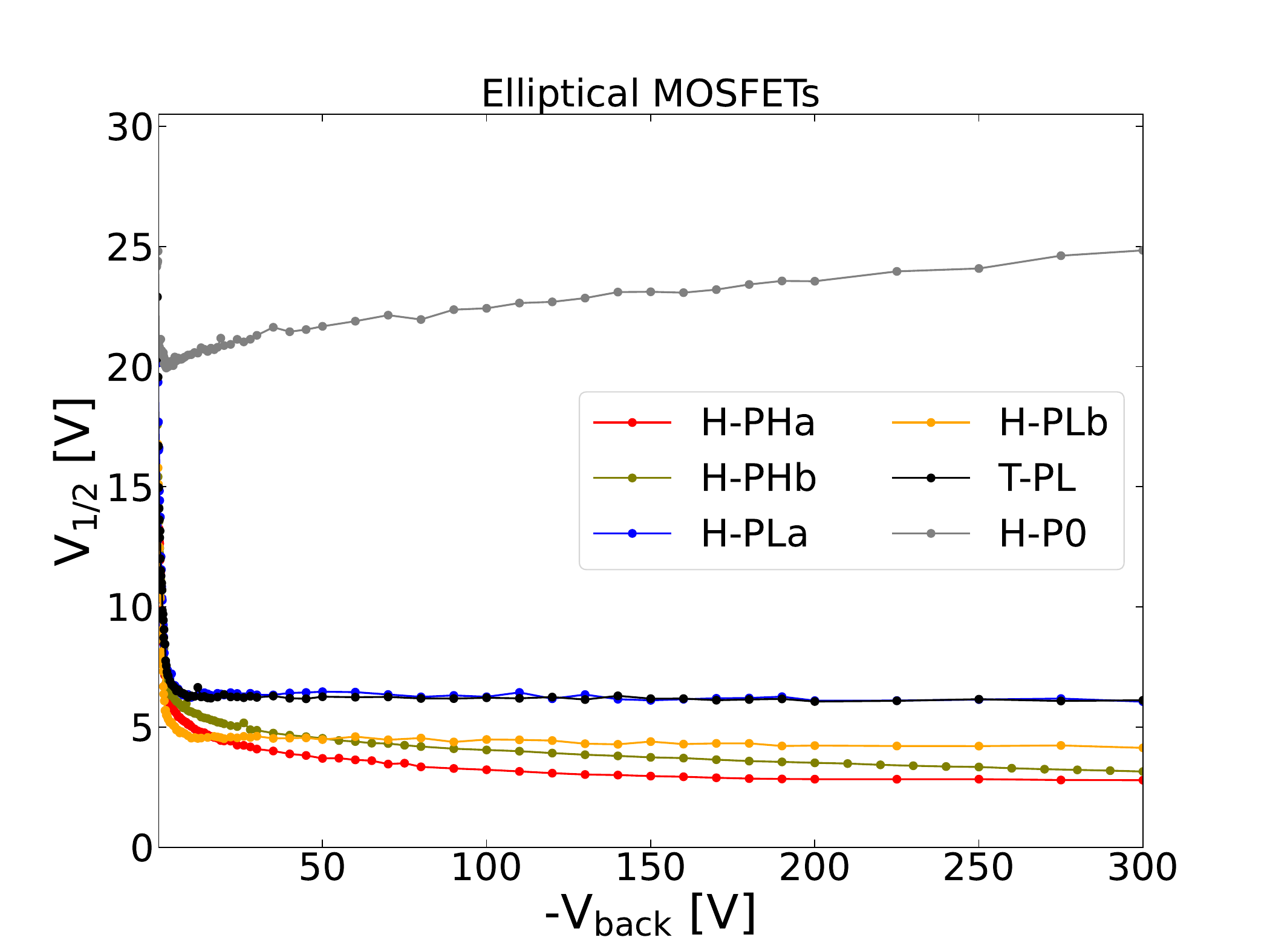}}
    \caption{
    Values of \Vh as a function of \Vback for circular (left) and elliptical (right) MOSFETs. Each color corresponds to a different p-stop configuration for the respective MOSFET geometry.
    }
    \label{fig:v1by2vsVbias}
\end{figure}

A similar comparison can be made for the extracted values of the ${\mu}_0$ and {\Vh} parameters as a function of {\Vback}, shown in Figs.~\ref{fig:mu0vsVbias} and~\ref{fig:v1by2vsVbias}, respectively.
For devices without p-stop implants, ${\mu}_0$ remains largely independent of {\Vback}.
Conversely, in structures including p-stop implants, ${\mu}_0$ first increases promptly with \Vmback and then saturates. 
Notably, the extracted electron mobility is consistently higher in the presence of p-stop implants, as the p-stop doping modifies the transverse electric field near the interface and introduces non-uniformities that can influence surface scattering~\cite{Nayfeh2003InfluenceOH,337449}.
The higher the p-stop doping concentration, the larger the mobility plateau value.
The parameter \Vh shows an approximately opposite trend to ${\mu}_0$. 
For both wafer types and MOSFET geometries, \Vh decreases with \Vback before reaching a plateau for structures containing p-stop implants, whereas for structures without p-stops, \Vh exhibits a slow increase with {\Vback}.

The initial steep rise in \Vth and the concurrent increase in ${\mu}_0$ are driven by the gradual depletion of the p-stop implant as {\Vmback} increases. This depletion modifies the effective space charge density at the Si-{\SiO} interface. Once the p-stop implant is fully depleted, the potential at the interface stabilizes, leading to the observed saturation or a significantly reduced rate of change in all three parameters at higher backside voltages. This characteristic transition point serves as an indicator of the total p-stop doping concentration and its effectiveness in isolating adjacent electrodes.

Both geometries exhibit the same qualitative dependence of the threshold voltage and mobility parameters on backside bias. The larger spread observed among the elliptical structures reflects the wider range of p-stop doping concentrations and the inclusion of two wafer types, whereas the circular devices are available only on wafer T. The present measurements do not allow either geometry to be identified as superior.

\subsection{Extraction of p-stop doping profiles}
\label{sec:profile}

The results presented in the previous section demonstrate that the extracted mobility parameters are highly sensitive to the details of the p-stop configuration, including the number of implants and their doping concentrations.  
Following the methodology described in Ref.~\cite{Fretwurst:2017dvs}, the doping profile in the inter-electrode region can be extracted from the dependence of the threshold voltage on the backside voltage.
In this section, this dependence is utilized to derive the effective doping profiles of the investigated structures.

The analytical formulae to determine the depletion depth ($x$) and the doping concentration ($p$) from the {\Vth}-vs-{\Vback} variation are given in Eqs.~\eqref{Eq:doping_depth} and~\eqref{Eq:doping_density}~\cite{SHANNON19711099,buehler1977dopant}.
\begin{equation}
x = \frac{\epSi}{\Cox} \diff{\Vback}{\Vth}
\label{Eq:doping_depth}
\end{equation}
\begin{equation}
p = \frac{{\Cox}^2}{\qzero \epSi} \left( \diff[2]{\Vback}{\Vth} \right)^{-1}
\label{Eq:doping_density}
\end{equation}
These equations allow for the reconstruction of the doping profile by relating the change in {\Vth} to the incremental depletion of the underlying space-charge region under the influence of the backplane bias.

The measured {\ISD} values are subject to experimental noise, which propagates through the numerical fitting procedure and amplifies the fluctuations in mobility parameters, including {\Vth}.
In particular, since Eq.~\eqref{Eq:doping_density} involves the inverse of the second derivative, small fluctuations in \Vth can lead to large artificial spikes in $p$.
To mitigate this effect, a smoothing procedure is applied to the {\Vth}-{\Vback} data prior to numerical differentiation, as outlined in the following. 
To compute the first derivative in Eq.~\eqref{Eq:doping_depth} at a given value of {\Vback}, two neighboring data points above and two below the selected {\Vback} are considered. 
In total, five points are fitted with a second-order polynomial. 
At the boundaries of the measurement range, only the available neighboring points are used in the polynomial fit.
The first derivative is then determined as the average of the numerical derivatives calculated using the {\Vth} values obtained analytically from the fitted polynomial at the selected {\Vback} value and at the adjacent points above and below it. 
It is verified that the first derivative obtained analytically from the fitted polynomial agrees well with the numerical differentiation results, with differences reaching at most a few percent.
Alternative configurations, varying both the number of points included in the derivative calculation and the order of the polynomial, have also been tested. 
The five-point second-order polynomial fit described above is found to provide stable results while keeping the number of parameters and data points minimal.  
Variations in the number of points used for the polynomial fit have been found to change the reconstructed depletion depth $x$ by up to 5\%.

\begin{figure}[hbtp]
    \centering
    {\includegraphics[width=0.495\textwidth]{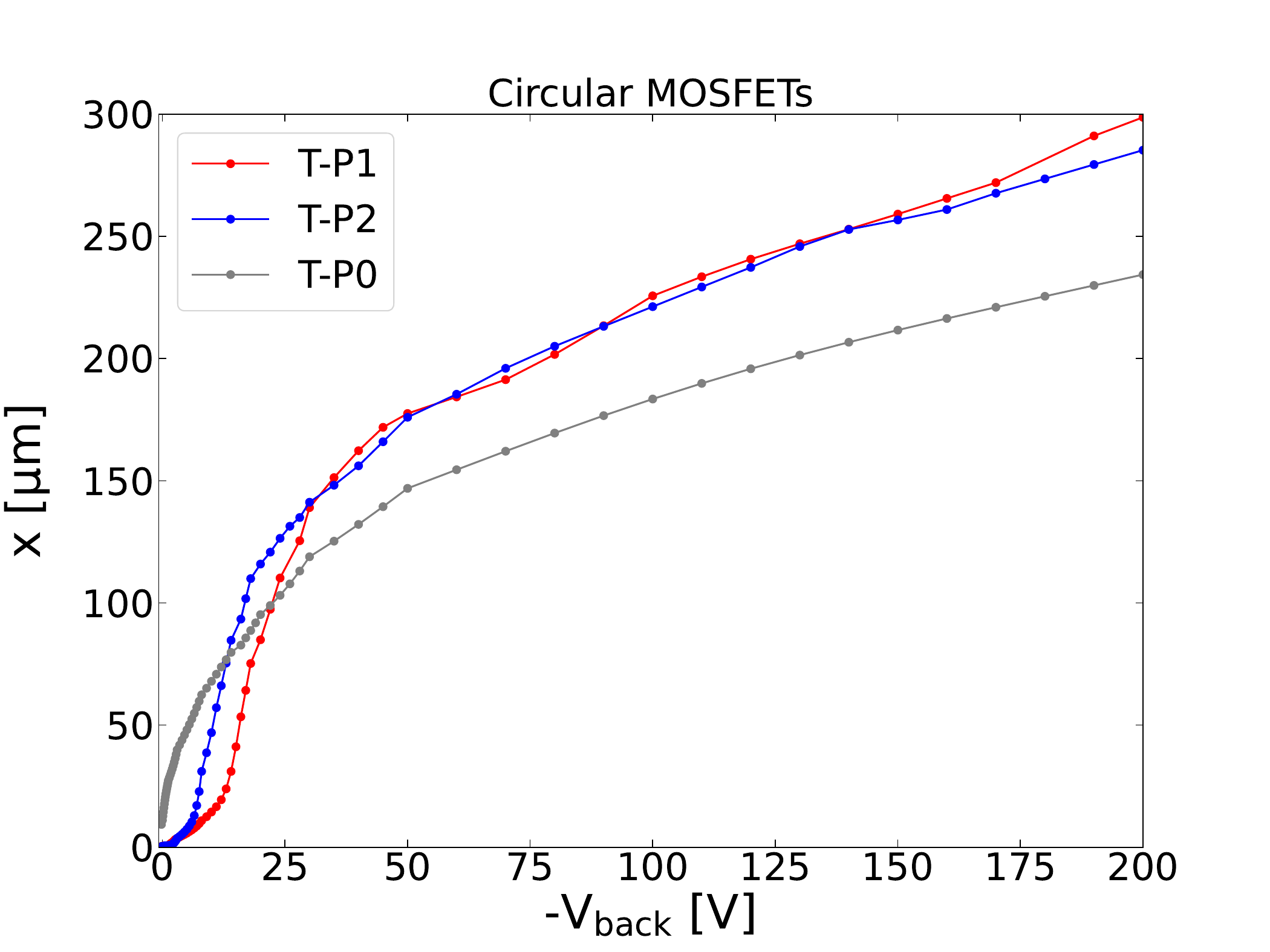}}
    {\includegraphics[width=0.495\textwidth]{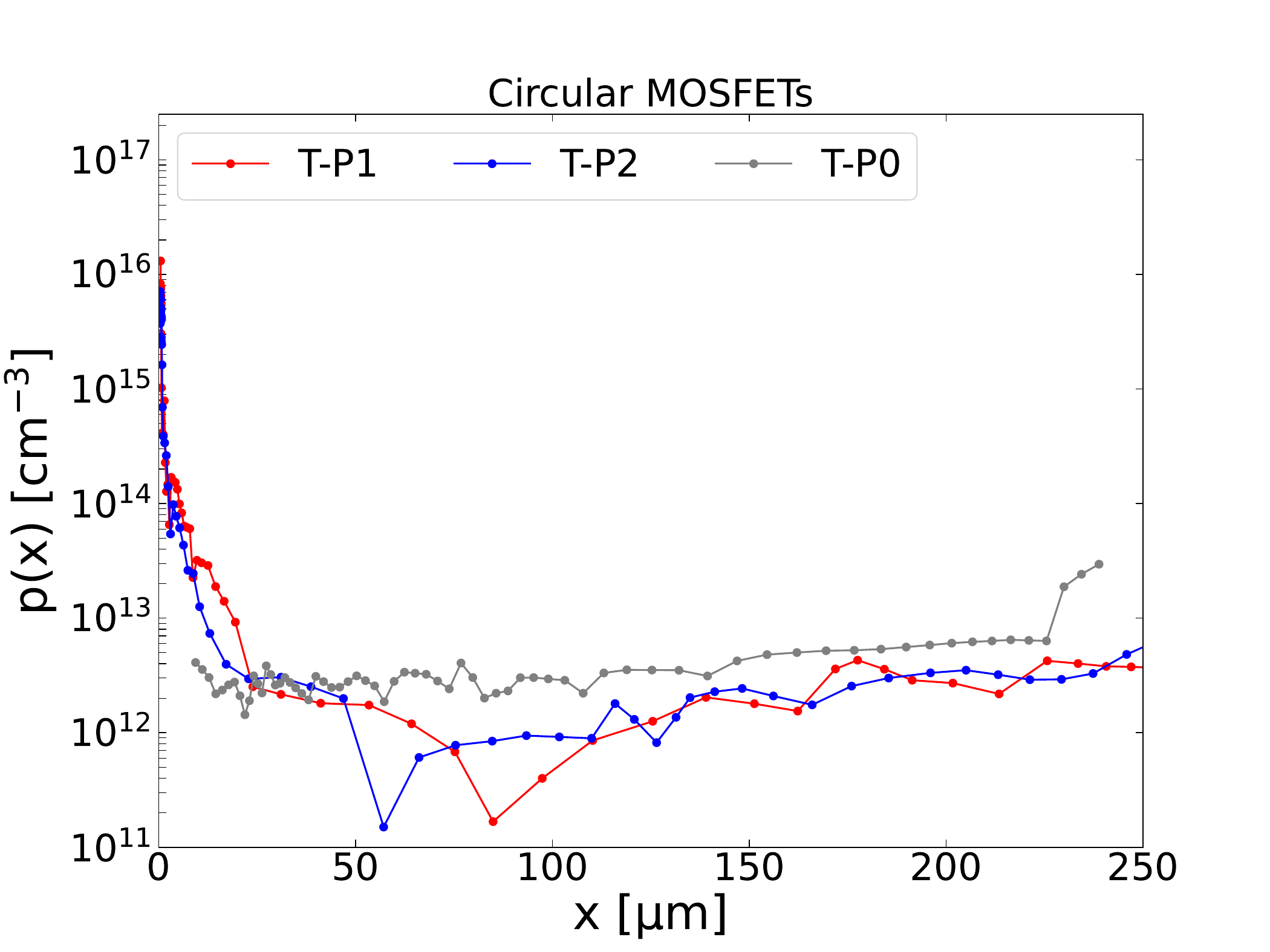}}
    {\includegraphics[width=0.495\textwidth]{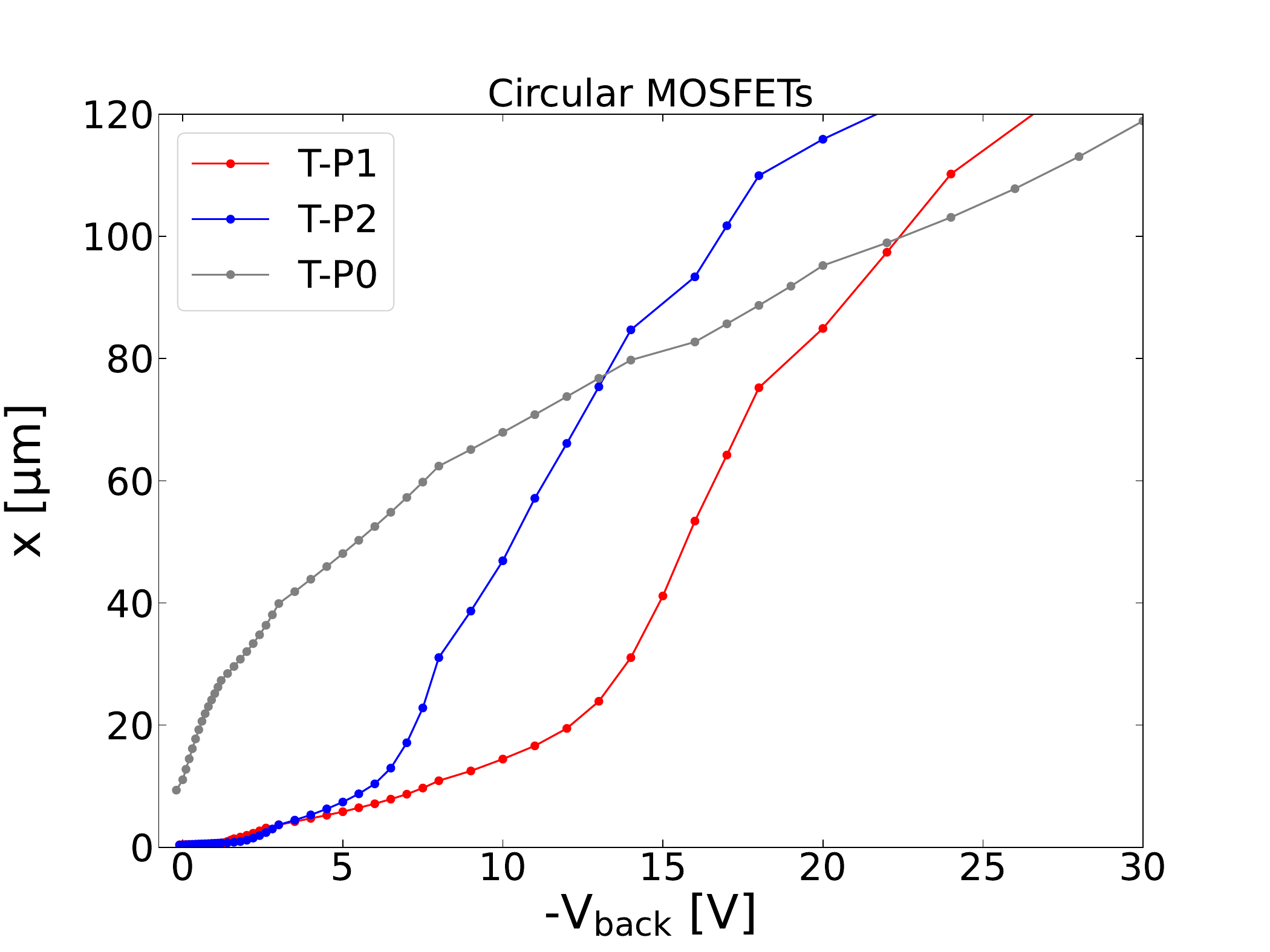}}
    {\includegraphics[width=0.495\textwidth]{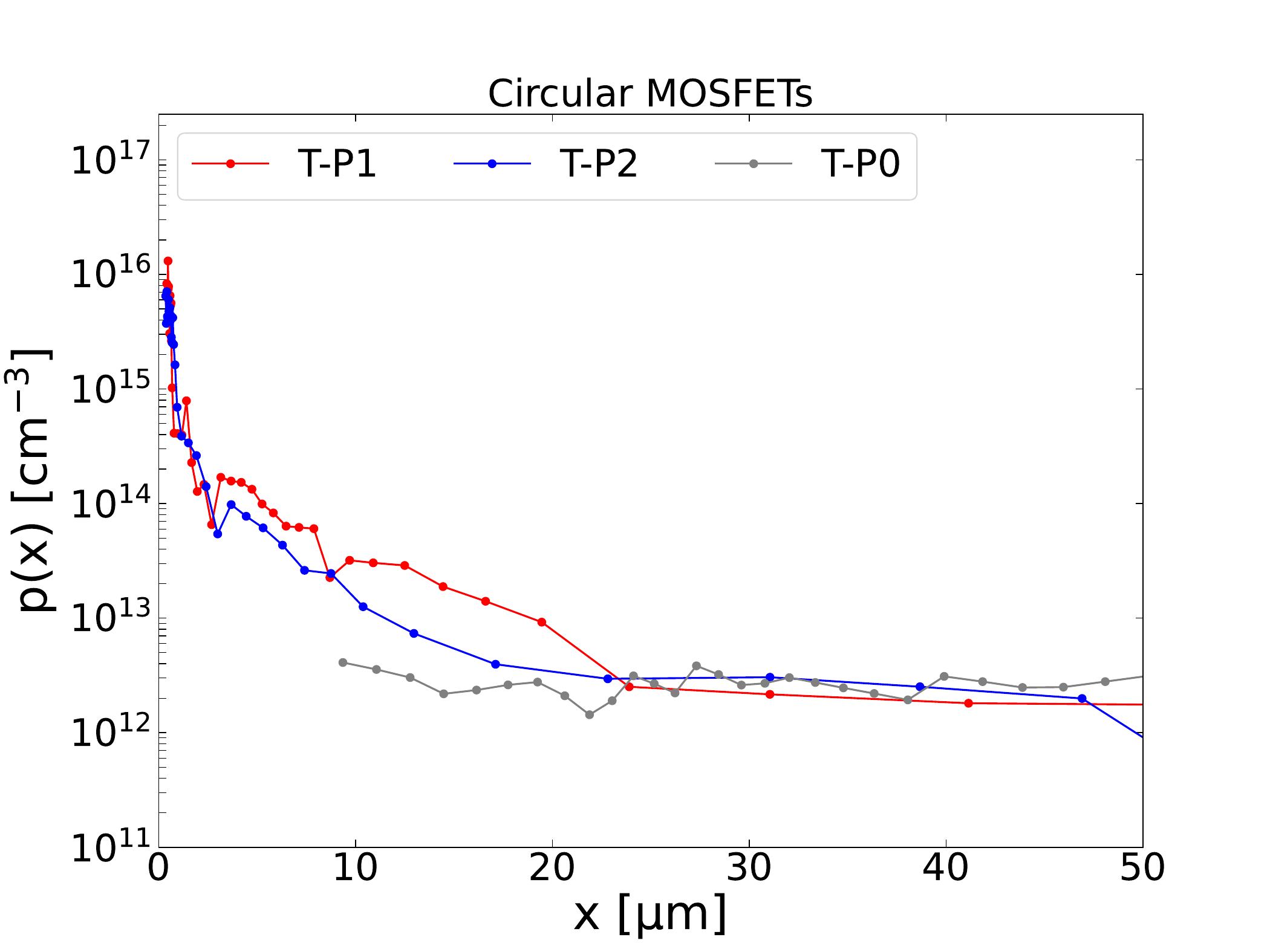}}
    {\includegraphics[width=0.495\textwidth]{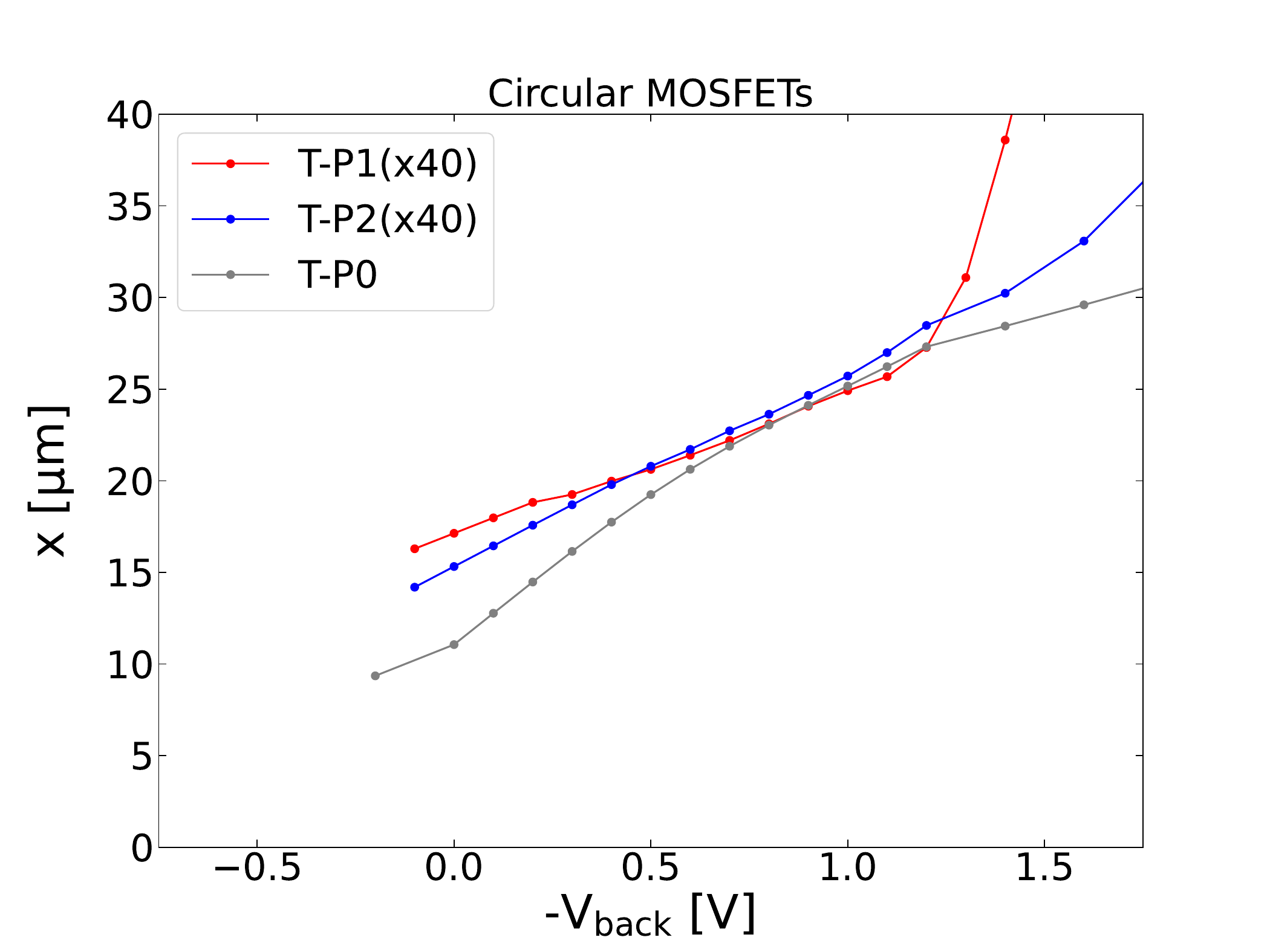}}
    {\includegraphics[width=0.495\textwidth]{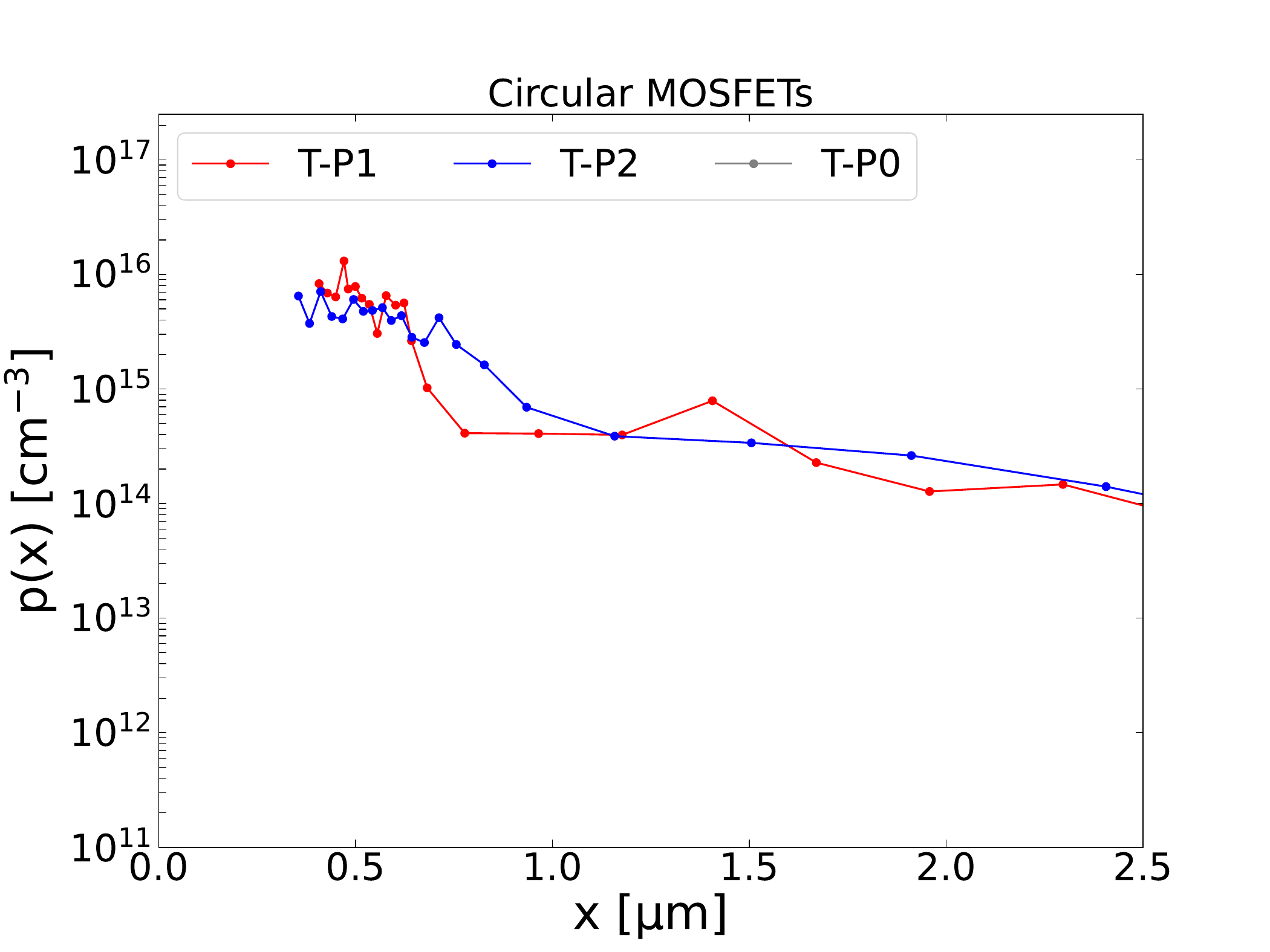}}
    \caption{
    Depletion depth as a function of \Vback (left) and doping concentration as a function of depletion depth (right) for the circular MOSFET structures on wafer T. Colors indicate different p-stop configurations. Different rows correspond to large (top), medium (middle), and small (bottom) ranges of {\Vback} (left) and depletion depth (right). 
    }
    \label{fig:doping_profile_HGC_trackerwafer}
\end{figure}

The variation of the resulting depletion depth, $x$, with \Vback for circular MOSFETs on wafer T is shown in Fig.~\ref{fig:doping_profile_HGC_trackerwafer} (left).
The structures with p-stop implants share a common three-stage trend: an initial regime of slow growth of $x$ for very small values of {\Vmback}, followed by a steep increase, and finally a slow rise at higher bias. 
In contrast, the structure without p-stops, T-P0, exhibits a smooth and monotonic increase of $x$ with {\Vmback} over the full bias range as expected for the depletion of a standard p-n junction.
The distinct behavior, originating from the specific surface charge distribution, is further emphasized by zooming in on a \Vback region in the middle row of Fig.~\ref{fig:doping_profile_HGC_trackerwafer}.
Because the p-stop implants introduce localized regions of higher acceptor concentration near the surface, the high density of negative space charge effectively screens the Si-{\SiO} interface, and hence the gate region, from the electric field resulting from the backside.  
Consequently, at very small \Vmback values, as shown in the bottom row of Fig.~\ref{fig:doping_profile_HGC_trackerwafer}, the depletion depth in p-stop structures is nearly an order of magnitude smaller than that in T-P0. Thus, the $x$ values for the p-stop structures are scaled by a factor of 40 to compare numbers visually in this regime.
When the \Vmback value is sufficiently large to compensate for the integrated charge of p-stop implants, the depletion region ``breaks through'' the potential barrier and starts to expand into the bulk. 

To obtain $p$, a cubic spline interpolation of the $x$-vs-{\Vth} data points is used to compute the second derivative appearing in the denominator of Eq.~\eqref{Eq:doping_density}. The smoothing procedure used to calculate $x$, as described above, also affects the reconstructed $p$; variations in parameters involved in the procedure, such as the number of points used for the polynomial fit, have been found to alter the doping concentration $p$ by up to 15\%.
The reconstructed doping concentration, $p$, as a function of $x$ is shown in Fig.~\ref{fig:doping_profile_HGC_trackerwafer} (right) for the same circular structures on wafer T.
For T-P0, $p$ remains roughly constant as expected for a uniformly doped bulk. 
The derived doping concentration is also compatible with the bulk doping level of approximately $4 \times 10^{12}\;\mathrm{cm}^{-3}$, as determined from capacitance–voltage measurements of diodes~\cite{CMSTrackerGroup:2021unz}.
On the other hand, the test structures with one or two p-stop implants show a characteristic trend of high $p$ values at small $x$ near the Si-{\SiO} interface. The  $p$ values decrease rapidly with depth and approach the bulk doping concentration beyond approximately 20\;{\mum}.
This gradient reflects the expected Gaussian-like profile resulting from the ion implantation process.
An estimate of the doping concentration of the p-stop implant is obtained by focusing on very small $x$ values shown in the bottom row of Fig.~\ref{fig:doping_profile_HGC_trackerwafer}.
The maximum value of $p$ is close to $10^{16}$\;cm$^{-3}$, consistent with estimates from  spreading resistance profiling measurements~\cite{CMS:2020jbh}. 
The doping concentration decreases by more than an order of magnitude from its maximum value within 1\;$\upmu$m, consistent with the expected thermal diffusion during sensor fabrication.

\begin{figure}[hbtp]
    \centering
    {\includegraphics[width=0.495\textwidth]{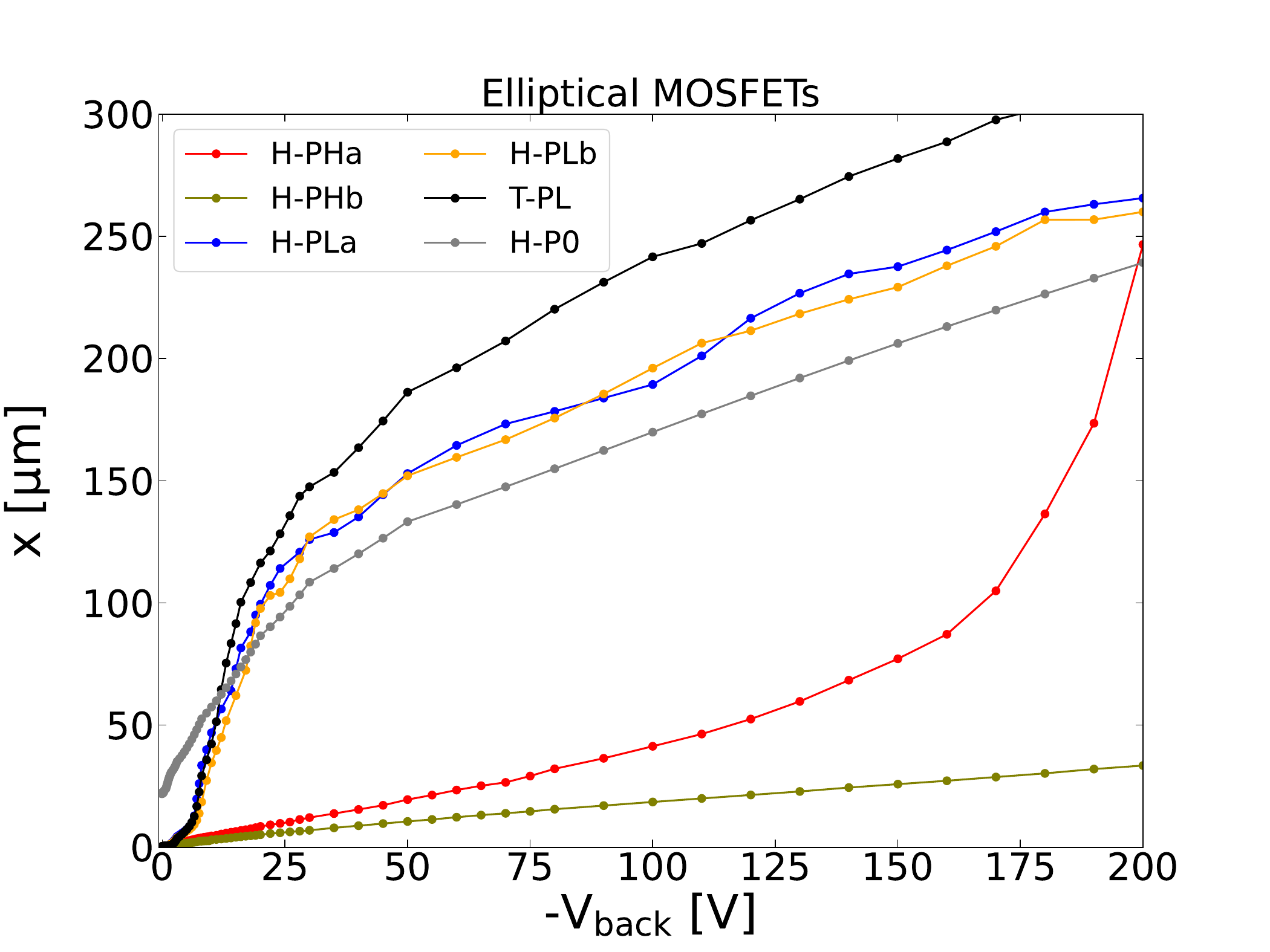}}
	{\includegraphics[width=0.495\textwidth]{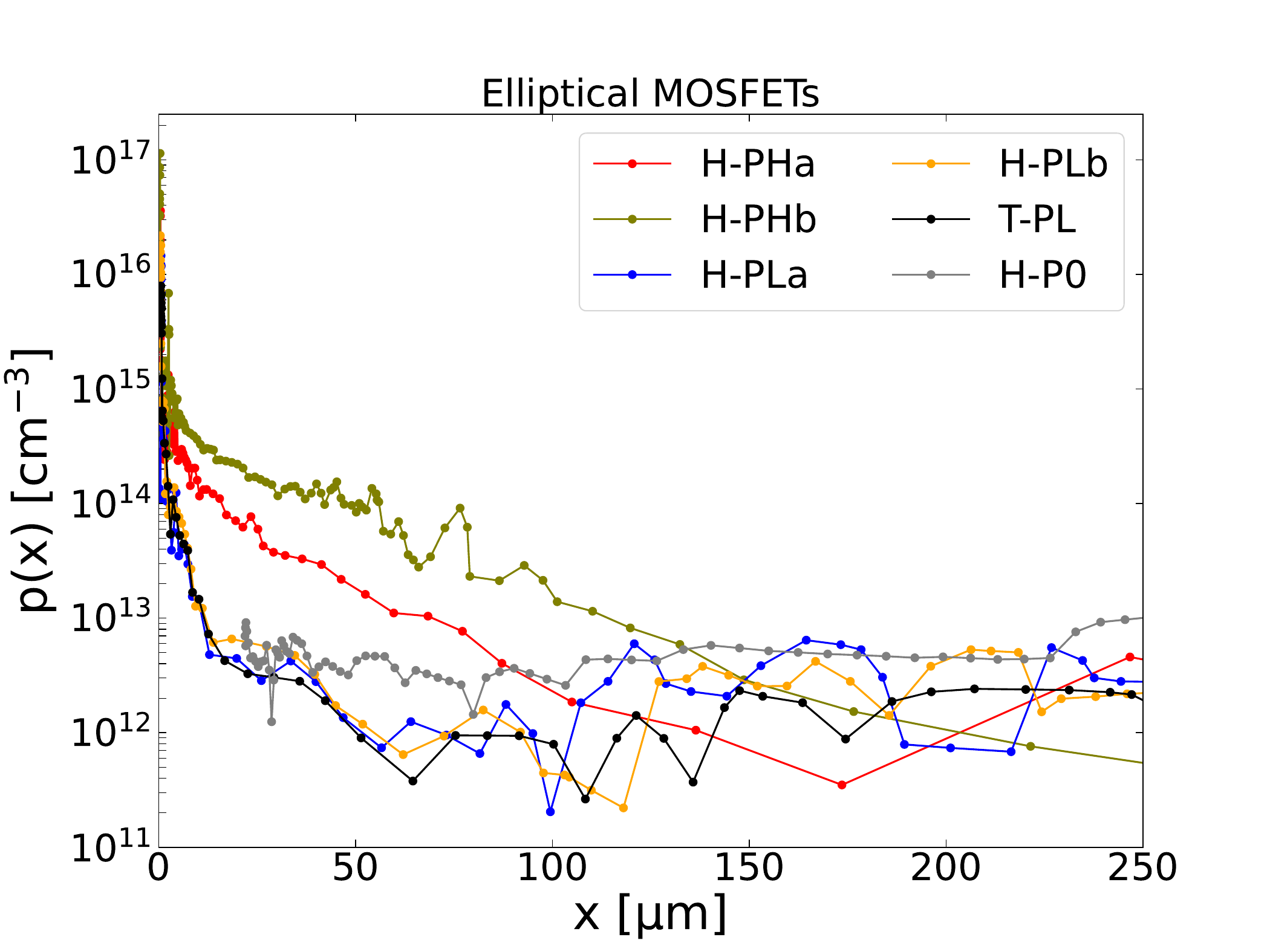}}    
    {\includegraphics[width=0.495\textwidth]{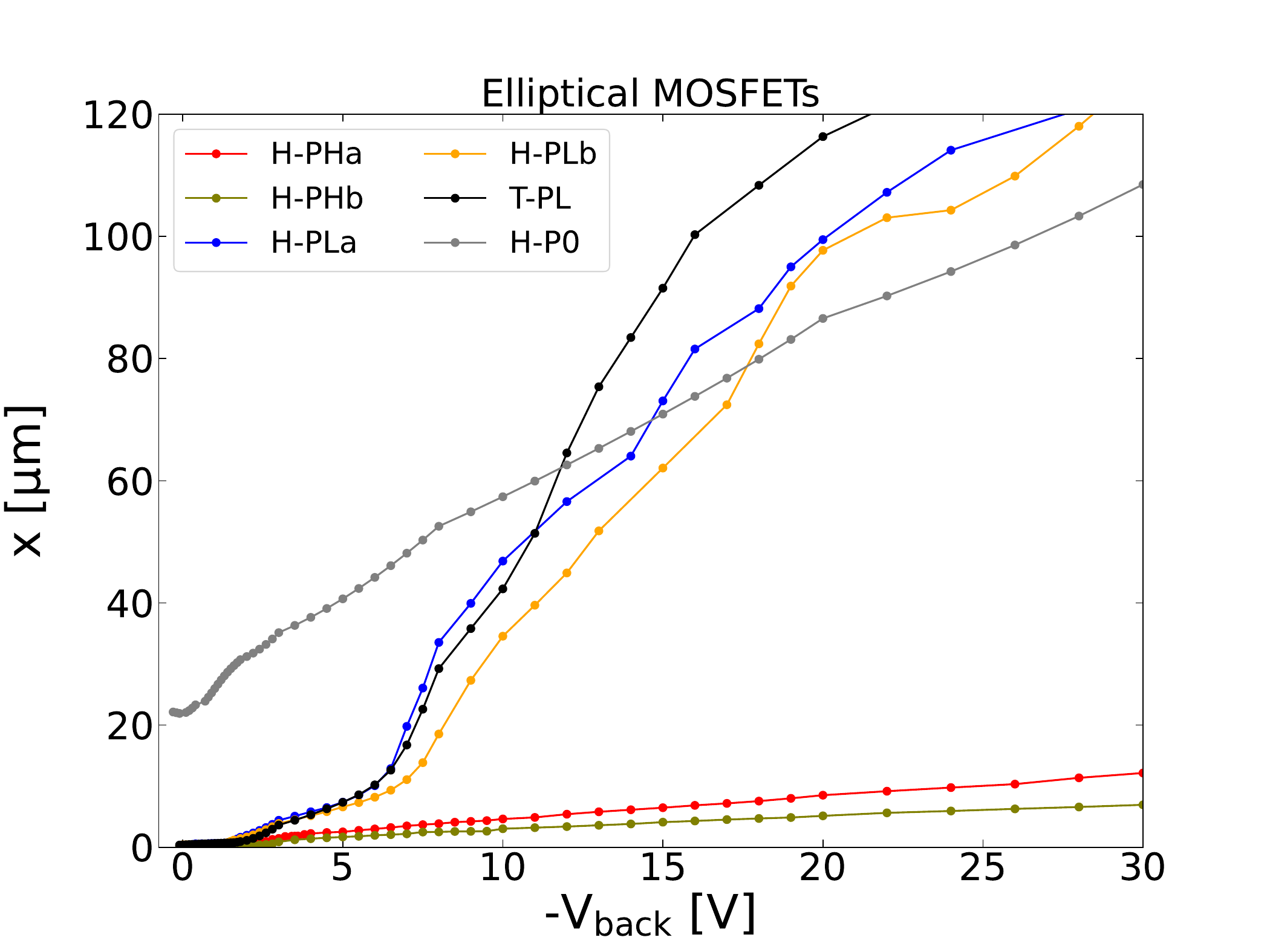}}
    {\includegraphics[width=0.495\textwidth]{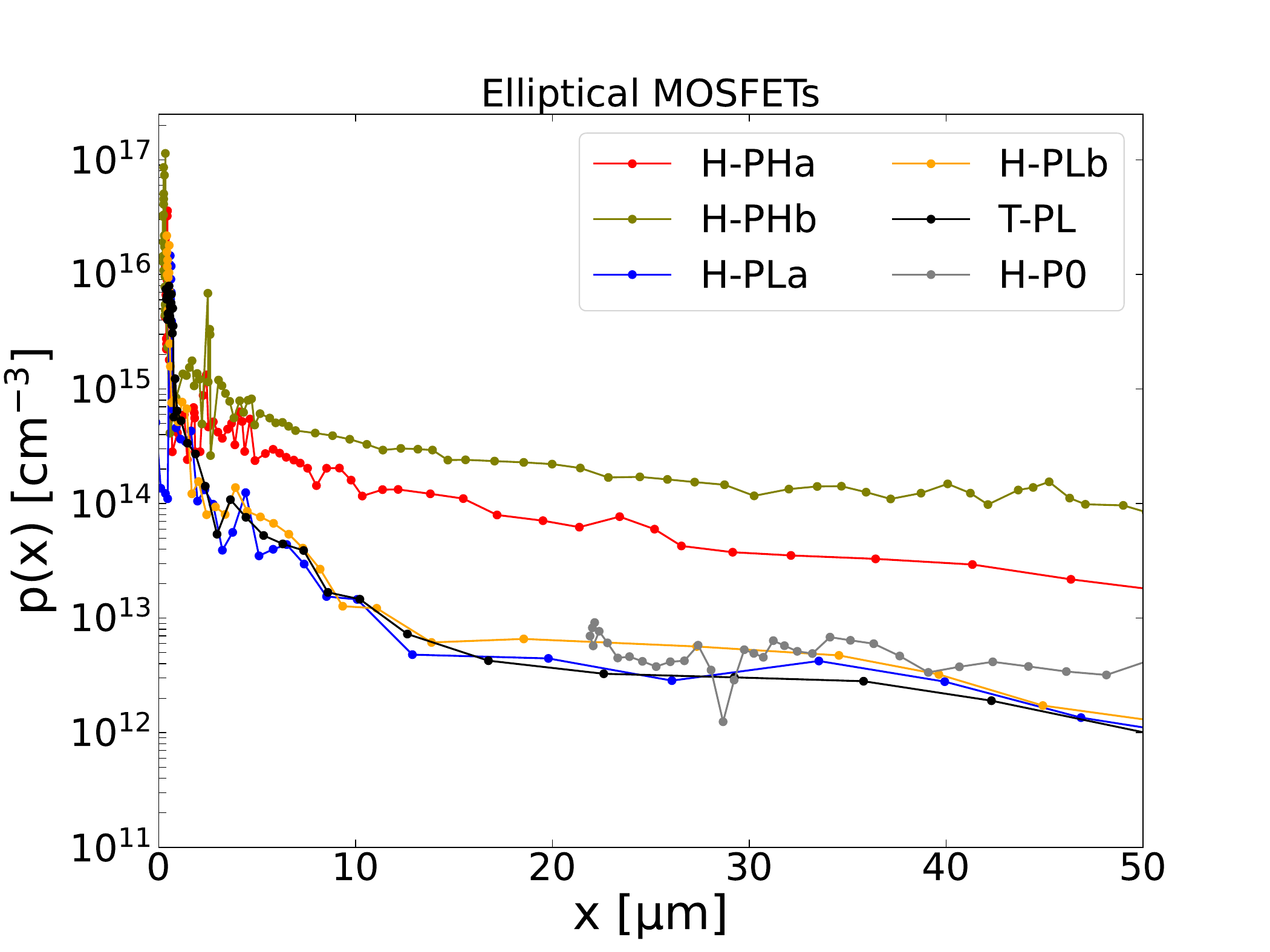}}
    {\includegraphics[width=0.495\textwidth]{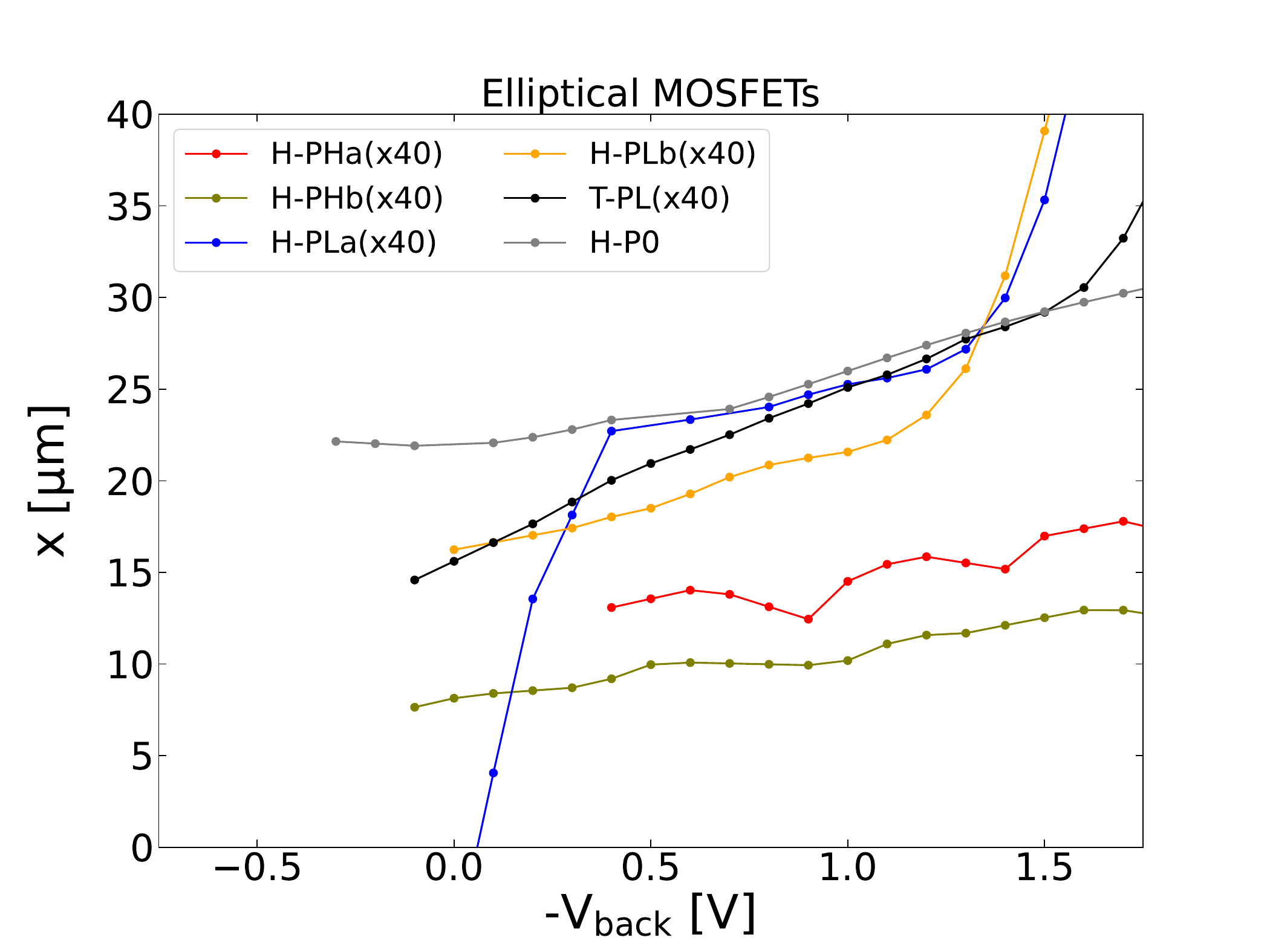}}
    {\includegraphics[width=0.495\textwidth]{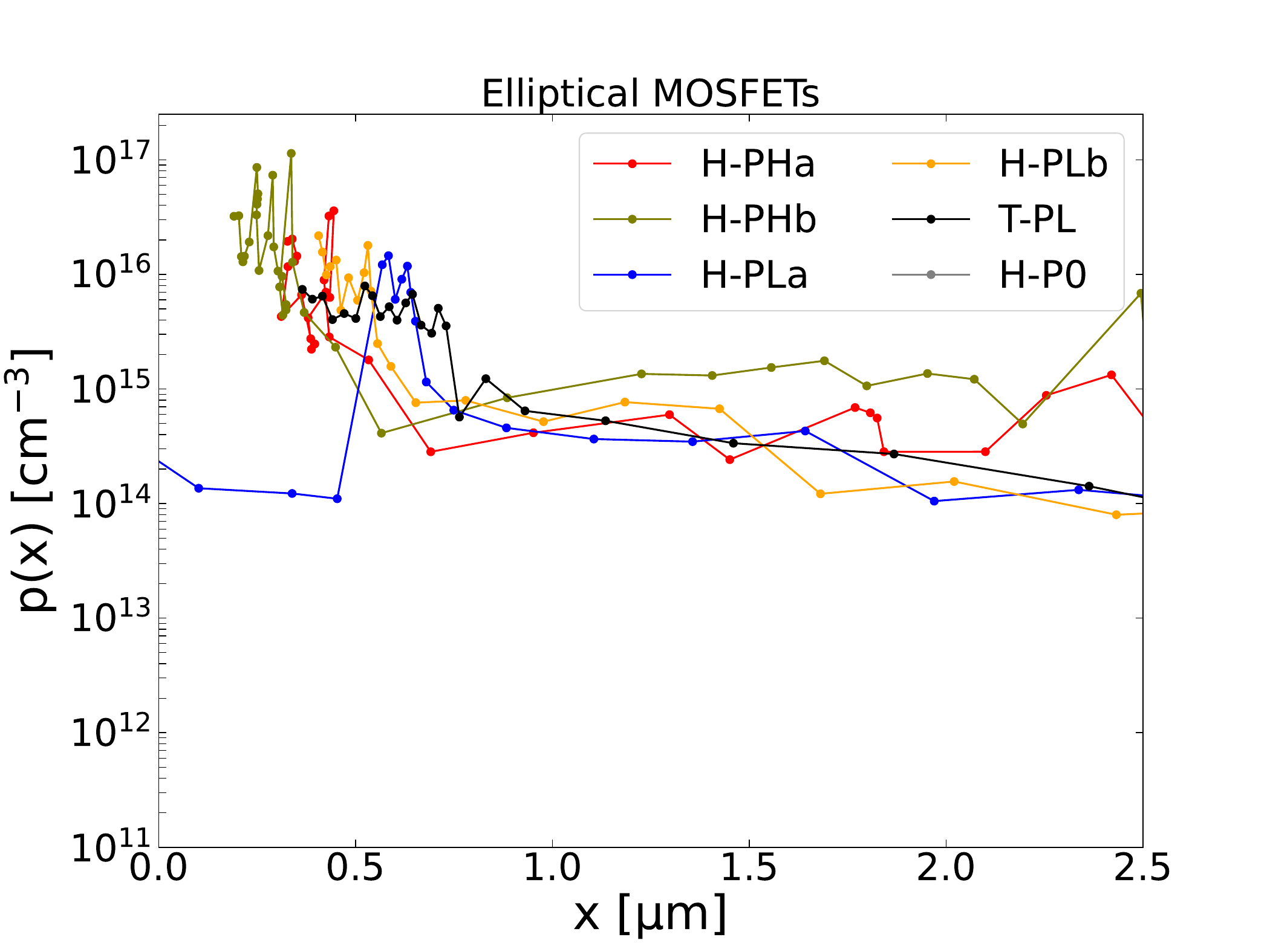}}
    \caption{
    Depletion depth as a function of \Vback (left) and doping concentration as a function of depletion depth (right) for the elliptical MOSFETs on wafers T and H. Colors indicate different p-stop configurations. Different rows correspond to large (top), medium (middle), and small (bottom) ranges of {\Vback} (left) and depletion depth (right). 
    }
    \label{fig:doping_profile_HGC_hgcalwafer}
\end{figure}

Similar results for the elliptical test structures on wafers T and H are presented in Fig.~\ref{fig:doping_profile_HGC_hgcalwafer}.
The low-doping p-stop variants on wafer H, H-PLa and H-PLb, exhibit a similar evolution of the depletion depth  with {\Vback}.
While the elliptical structure including p-stop implantation on wafer T shows a similar trend at small {\Vmback}, it reaches larger depletion depths for higher {\Vmback} values compared to its low-doping H-wafer counterparts.
In contrast, the high-doping H-wafer MOSFETs, i.e., H-PHa and H-PHb, show a weaker dependence of $x$ on \Vback than H-PLa, H-PLb, and T-PL.
This is particularly evident at very small {\Vmback}, as shown in the bottom row of Fig.~\ref{fig:doping_profile_HGC_hgcalwafer}, where the depletion depths for H-PHa and H-PHb are smaller by factors of approximately 2 and 4, respectively, than that for T-PL. This indicates that the higher integrated charge in these p-stops provides stronger screening of the external electric field.
While the doping profiles of the low-doping variants are similar across wafers, those of H-PHa and H-PHb have much longer tails, corresponding to a slower decrease of $p$ as a function of $x$.
The higher the p-stop doping concentration, the greater the depletion depth required for $p$ to approach the bulk doping concentration.

\section{Conclusion}
\label{sec:conclusion}

In this paper, we have investigated the electrical characteristics of different MOSFET test structures fabricated on p-type silicon wafers and evaluated their potential for characterizing p-stop implants in inter-electrode regions. 
By analyzing the transfer characteristics as a function of the backside bias voltage, we have demonstrated a methodology for extracting device parameters, including the threshold voltage and mobility parameters.

A strong correlation between the p-stop implant configuration and the bias dependence of the MOSFET parameters is observed. 
The presence of p-stops induces a characteristic steep increase in the threshold voltage at low bias voltages, followed by saturation once the implant is fully depleted.
This behavior is leveraged to reconstruct the depth-dependent doping concentration in the inter-electrode region with sub-{\mum} resolution near the surface. 
The extracted peak doping concentrations, approaching $10^{16}$\;cm$^{-3}$, and the decrease in doping concentration by more than an order of magnitude within approximately 1\;{\mum} are consistent with resistivity measurements and fabrication specifications.
The results therefore provide quantitative information on the depletion behavior and electric-field screening associated with p-stop implants.

Furthermore, the study establishes that structures with higher p-stop doping more strongly constrain the depletion region and require substantially higher backside voltages to reach a depletion regime comparable to that observed in lower-doped structures. 
These results demonstrate that MOSFET test structures provide a non-destructive diagnostic tool for monitoring p-stop doping consistency and sensor isolation properties during the production of high-granularity silicon detectors, such as those developed for the High-Luminosity LHC.

\acknowledgments

We thank the Tracker and High-Granularity Calorimeter groups of the CMS Collaboration for allowing us to use the test structures produced for sensor qualification. 
We also thank Andreas Bauer and Konstantinos Damanakis for technical assistance during the project. 

\bibliography{MOSFET_Vienna}
\bibliographystyle{JHEP}

\end{document}